\documentclass[a4paper,preprintnumbers,floatfix,twocolumn,aps,prb,unsortedaddress,superscriptaddress]{revtex4-2}

\usepackage{miller}
\usepackage{graphicx}
\usepackage{dcolumn}
\usepackage{bm}
\usepackage{subcaption}
\usepackage{textgreek}
\usepackage{booktabs}

\usepackage[paperwidth=210mm,paperheight=297mm,centering,hmargin=1.7cm,vmargin=2.5cm]{geometry}
\usepackage{url}

\usepackage[colorlinks=true, allcolors={magenta}]{hyperref}

\begin{document}

\title{Efficient atomistic simulations of radiation damage in W and W-Mo using machine-learning potentials}

\author{Mikko Koskenniemi}
\email{Corresponding author, mikko.a.koskenniemi@helsinki.fi}
\affiliation{Department of Physics, P.O. Box 43, FI-00014 University of Helsinki, Finland}
\author{Jesper Byggmästar}
\affiliation{Department of Physics, P.O. Box 43, FI-00014 University of Helsinki, Finland}
\author{Kai Nordlund}
\affiliation{Department of Physics, P.O. Box 43, FI-00014 University of Helsinki, Finland}
\author{Flyura Djurabekova}
\affiliation{Department of Physics, P.O. Box 43, FI-00014 University of Helsinki, Finland}
\affiliation{Helsinki Institute of Physics, Helsinki, Finland}
\date{\today}

\begin{abstract}
    The Gaussian approximation potential (GAP) is an accurate machine-learning interatomic potential that was recently extended to include the description of radiation effects. 
    In this study, we seek to validate a faster version of GAP, known as tabulated GAP (tabGAP), by modelling primary radiation damage in 50-50 W-Mo alloys and pure W using classical molecular dynamics.
    We find that W-Mo exhibits a similar number of surviving defects as in pure W. We also observe W-Mo to possess both more efficient recombination of defects produced during the initial phase of the cascades, and in some cases, unlike pure W,  recombination of all defects after the cascades cooled down. 
    Furthermore, we observe that the tabGAP is two orders of magnitude faster than GAP, but produces a comparable number of surviving defects and cluster sizes. A small difference is noted in the fraction of interstitials that are bound into clusters. 
\end{abstract}

\maketitle
\section{Introduction}

Nuclear energy is an integral part of modern society; nuclear fuels are millions of times more energy-dense than chemical ones, such as oil. Moreover, they release no greenhouse gases. The materials in nuclear reactors are exposed to intense irradiation, and the understanding of the consequences of this process on the durability and reliability of the materials is vital not only for existing power plants but more so for future fusion and next-generation fission reactors~\cite{zinkle_designing_2014}. This motivates the search for new radiation-tolerant materials. Tungsten-based high-entropy alloys (HEA) are a class of materials that show promising resilience to radiation~\cite{W_high-entropy_alloys}, making them particularly interesting in the field of nuclear energy applications.

Molecular dynamics \cite{Allen-Tildesley} (MD) is a widely used  method to study how materials respond to radiation and gives insight into atomic-scale phenomena and their underlying mechanisms that are inaccessible by experimental means~\cite{Nor18}. Considering specifically W-based alloys,
Qiu et al.~\cite{W-Ta_cascades} found, by running collision-cascade simulations, that alloying Ta with W can decrease the size of dislocation loops, whilst retaining comparable defect production to W. Moreover, cascade simulations have shown Mo-based complex concentrated alloys to fare well under radiation~\cite{Mo-based_CCAs}. However, the effects of collision cascades in W-based alloys are still fairly poorly understood.

Interatomic potentials that describe the nature of atom interactions within the modelled material are essential for the validity and accuracy of simulation results. 
However, analytical potentials (potentials that have a fixed mathematical form, comprising only a few parameters) struggle to accurately describe more than a handful of phenomena, fundamentally restricting the use of their applications.
Recently, a new approach to the development of interatomic potentials based on machine-learning (ML) algorithms was proposed \cite{GAP_paper,behler_generalized_2007}. Since the training database is generated from consistent density functional theory (DFT) calculations, some of the ML potentials excel at describing a multitude of different phenomena, giving more accurate results than their analytical counterparts~\cite{behler_generalized_2007,GAP_paper, machine-learned_potential_in_W}.

The Gaussian approximation potential (GAP)~\cite{GAP_paper} is a popular machine-learning potential, which has been proven to give results that are on par with quantum-mechanical simulation methods, and is capable of successfully describing a diverse range of phenomena~\cite{GAP_result_accuracy1, GAP_result_accuracy2}.
GAP also reaps the benefits of classical potentials, being capable of simulating systems that are at least thousands of times larger than in quantum-mechanical methods. Despite this, GAP is still excruciatingly slow when put up against its traditional, analytical counterparts, such as the embedded atom method (EAM) potentials. In an attempt to retain the excellent array of properties of GAP, whilst making it faster to compute, the tabulated GAP (tabGAP) formalism was created~\cite{tabGAP,byggmastar_simple_2022}.

The key feature of tabGAP is using only low-dimensional descriptor terms, omitting terms like the \textit{Smooth Overlap of Atomic Positions (SOAP)} term~\cite{SOAP}, which is a vector in a space of hundreds or even thousands of dimensions for multi-component materials.
The low-dimensional terms enable tabGAP to circumvent the exhausting machine-learning prediction of GAP when computing atomic energies by using tabulation.
Tabulation involves pre-computing the GAP energy predictions and mapping them onto low-dimensional grids. After tabulation, the resulting data grid can be used in conventional spline interpolation methods during simulations, which makes tabGAP faster. 
Perhaps even more importantly, the low-dimensional terms of tabGAP make it easier to develop for many-element materials like HEAs because they need less training data than terms like SOAP~\cite{byggmastar_simple_2022}. Therefore, tabGAP could act as a gateway to efficient, and accurate, studies of exotic multi-component materials. 
\\
\\
In the present study, we test the tabGAP developed in Ref.~\cite{tabGAP}, which was developed for a W-based HEA, namely molybdenum-niobium-tantalum-vanadium-tungsten (Mo-Nb-Ta-V-W), by modelling radiation effects.
To compare the performance of tabGAP to other types of interatomic potentials in MD simulations, we choose to model 50-50 W--Mo alloys. We note that the high activation of Mo under neutron irradiation limits the use of this particular alloy for fusion applications; however, it could be used in small amounts e.g. in fusion reactor diagnostics, and in non-fusion applications where neutron activation is not an issue.
Our choice is motivated by the existence of both a GAP and EAM for W-Mo~\cite{W-Mo_GAP,CHEN2020152020}. 
Additionally, the results of this study give general insight into how 50-50 W-based refractory alloys behave.
Radiation damage in both 50-50 W-Mo alloys and pure W is modelled by the means of MD collision-cascade simulations using tabGAP, a SOAP-equipped GAP, and EAM. 
The simulation results are analysed for the number of surviving defects (point defects and their clusters).

\section{Methods}
\subsection{Software and potentials}

The simulations were run using the classical MD code, \textit{Large-scale Atomic/Molecular Massively Parallel Simulator} (LAMMPS)~\cite{LAMMPS} (\url{https://www.lammps.org/}). The QUIP code~\cite{GAP_paper} (\url{https://github.com/libAtoms/QUIP}) was used to enable the use of GAP with LAMMPS.
The \textit{Open Visualization Tool}~\cite{ovito} (OVITO) was used for both visualising simulation results and defect analysis using the Wigner-Seitz method. 
Dislocations were analysed using the \textit{Dislocation Extraction Algorithm}~\cite{DXA}.
The Python library Matplotlib~\cite{matplotlib} was used for plotting simulation data.
\\\\
Cascades were run using four potentials: the EAM potential developed for W-Mo in Ref.~\cite{CHEN2020152020} (hereafter referred to as \textit{W--Mo-EAM}), the Ackland-Thetford--Zhong-Nordlund (AT-ZN) EAM potential developed for W in Ref.~\cite{ackland_improved_1987,zhong_defect_1998},
the GAP developed in Ref.~\cite{W-Mo_GAP}, and the tabGAP developed in Ref.~\cite{tabGAP}. We chose the AT-ZN potential for pure W, for it is the most widely used potential for radiation damage simulations in W~\cite{W_subcascade,granberg_molecular_2021}. For example, it has shown good agreement with experiments and GAP at high doses~\cite{granberg_molecular_2021}, which makes a comparison to the other potentials useful.

All four potentials were developed to be applicable for the simulation of radiation effects, i.e. joined with corresponding repulsive potentials, such as the ZBL potential in EAM \cite{ZBL} and DMol \cite{Nor96d} in GAP and tabGAP, to enable a reasonable description of cascade development.

It is worth noting that the present tabGAP is fitted to a HEA dataset, whereas the GAP is fitted to a W--Mo dataset. In the HEA set, there are less data for the W--Mo system, which makes a direct comparison between GAP and tabGAP difficult. For more details about the development of the GAP and tabGAP, see Refs.~\cite{tabGAP} and~\cite{W-Mo_GAP}.
\subsection{Selection of the primary knock-on atom}

Following the practice in \cite{Nor96d}, cascades were initiated by giving one atom, the primary knock-on atom (PKA), a recoil of a given energy towards the centre of the simulation cell.
The PKAs were selected as follows. Firstly, we generate a random direction in three-dimensional space. Then, we define a point at a specific distance from the centre of the cell, in the aforementioned direction.
Finally, the atom closest to this point is given the recoil in the aforementioned direction, towards the cell centre, to initiate the cascade.
Higher recoil energies trigger more extensive cascades, hence the distance at which a PKA was selected, as well as the total number of atoms in the simulation cell, scale up with the recoil energy. These parameters are given in Table~\ref{tab:sim_params}.

In LAMMPS, the atoms within a simulation cell are labelled by identifiers (identification numbers). Since the same atomic structure for a given material was used for all potentials, for consistency, in the simulations with different potentials, we selected as a recoil the atom with the same identifier. We assigned it with the same velocity in the same direction. Although the cells relaxed in different potentials may slightly deviate from one another, these differences are sufficiently small for a statistically averaged quantitative comparison of defect formation in different potentials.
\\\\
It is worth noting that because the PKAs were selected in random directions, they may move in channelling directions (which offer the least resistance to movement), and a few cascades overlapped with the periodic boundaries, in spite of the sufficient size of the simulation cells. These simulations were discarded and the simulations were re-run with new PKAs. The aim of the PKA selection method is to minimise the direction-related bias in the results. 
Regardless, the present results are not completely free of directional bias, since the channelling directions were excluded from the analysis. 
However, the main purpose of the current paper, which is to compare the results of different interaction models, is unaffected by this, since the probability of crossing the boundaries is the same for all interaction models.  In fact, the number of failed simulations (where atoms enter the thermostatted border with at least 10-eV kinetic energy) was around five out of the 40 1- and 2-keV simulations, but only around two simulations for the rest of the energies (these energies gave rise to thermal spikes).

\begin{table}[h]
\begin{ruledtabular}
\centering
\caption{Simulation parameters. Here, 
$E_{\text{PKA}}$ is the initial kinetic energy of a PKA, $r_{\text{PKA}}$ is the distance from the PKA to the centre of the lattice, and $n_{\text{atoms}}$ is the number of atoms in the lattice.
}
\label{tab:sim_params}			
\begin{tabular}{ccc}
$E_{\text{PKA}}\, \left[ \text{keV} \right]$ & $r_{\text{PKA}}\,  \left[ \text{Å} \right]$ & $n_{\text{atoms}}$  \\ 
\midrule 
$ 1 $ & $ 15 $ & $ 31 ~ 250 $ \\
$ 2 $ & $ 15 $ & $ 54 ~ 000 $ \\
$ 5 $ & $ 20 $ & $ 159~014 $ \\
$ 10 $ & $ 30 $ & $ 332~750 $ \\
$ 20 $ & $ 40 $ & $ 686~000 $ \\
\end{tabular}
\end{ruledtabular}
\end{table}

\subsection{Simulation setup}

Collision cascade simulations were run for 50-50 W-Mo alloys, and pure W, both with the body-centred cubic (BCC) lattice structure. The atoms in the W-Mo alloys are randomly ordered.
Periodic boundary conditions were used in every simulation.

In W-Mo alloys, the cascades initiated by PKA with energies from $1$ to $20~\text{keV}$ were run using the EAM and tabGAP potentials, but only 1 to 5-keV cascades were run using GAP, due to its much higher computational cost (GAP is two orders of magnitude slower than the tabGAP we used and four orders of magnitude slower than the EAMs; see Tab.~\ref{tab:performance}).

In pure W, simulations were run using the AT-ZN EAM, the W-part of the W--Mo-EAM potential and the tabGAP to study stable defects and their clusters with PKA energies of 1 to 10~keV.
Only 1-keV cascade simulations were run in pure W with the GAP. For each PKA energy, statistics were collected over 40 simulations with different initial seeds for random-number generation, except for GAP 5~keV in W-Mo. In the latter case, only 25 simulations were run, again due to the prohibitively high computational cost of these simulations. Even the case of 25 simulations should be sufficient, as has been studied in Ref.~\cite{voskoboinikovOptimalSamplingMD2020}.

For consistency, in all applied potentials, we used cells of the same composition. Therefore, we relaxed the simulation cells with the corresponding potential before cascade simulations.
The relaxation was done by imposing a Nos\'e-Hoover thermostat and barostat to the cells \cite{Hoover_thermostat, Nose_thermostat}, and waiting for the pressure and volume of the cells to become stable. Cascade simulations started out at a temperature of $300~\text{K}$, and had a Nos\'e-Hoover thermostat applied to a $6$-Å thick shell at the boundary of the simulation cells, to cool the cell down to its initial temperature, which mimics the much larger bulk material surrounding the cascade region. During the cascade simulations, no pressure control was used.
The simulation time was chosen such that the final temperature is sufficiently close to the initial 300 K and the cascade-induced defect evolution has stopped. For each W-Mo simulation, it was $100~\text{ps}$, with the exception of $5$-keV GAP simulations, where the shortest simulation managed to run for about $71~\text{ps}$.
The shorter run-time was deemed a non-issue, as will be discussed in more detail in section \ref{sec:defects}.
For pure W, a shorter simulation time of $60~\text{ps}$ was sufficient.

Due to the nature of the cascade simulations, the initially-high kinetic energies of atoms (high velocities) decrease over time. For simulation efficiency, an adaptive time-step \cite{adaptive_time_step} was used. The magnitude of the adaptive time-step changes dynamically in response to atomic velocities, starting out small and ultimately reaching a fixed maximum value, which was chosen to be $3~\text{fs}$. 

In the MD simulations, electrons are not explicitly modelled, however, they do have a substantial role in energy dissipation for the collision energies involved in the cascades of this study \cite{NORDLUND2018450}. To emulate the energy loss due to electronic excitations of high-energy atoms, electronic stopping data were used to determine the magnitude of the electronic stopping power that the atoms experience at a given kinetic energy. A cut-off kinetic-energy threshold of $10~\text{eV}$ was used and the electronic stopping was applied to all atoms with kinetic energy higher than this.
The stopping power for the W-Mo alloys was generated using the SRIM-2013 code~\cite{SRIM-2013, SRIMbook}, while the stopping power for the pure W was the same as in the earlier work~\cite{W_subcascade}, generated with the ZBL-96 code~\cite{ZBL}. In the energy range of interest for the current study ($\leq$ 20~keV, well below the maximum in the electronic stopping power), the stopping power in both codes is based on the Lindhard stopping model~\cite{LSS}.
Hence, the possible difference in the stopping powers generated by both methods will have a negligible effect on defect formation.  

In addition to the cascade simulations, the mobility of interstitials was determined using tabGAP in both pure W, and 50-50 W-Mo cells. The simulation cells of perfect BCC lattices of 2~000 atoms with manually added 5–6 split-interstitials in random positions were modelled for 1 ns of simulated time using a 3-fs timestep. A single W simulation was run at $600~\text{K}$, and one W-Mo simulation at both $600~\text{K}$ and $1200~\text{K}\,$.
A thermostat and barostat were applied to these cells, making them $N\,P\,T$ ensembles. 
The purpose of these simulations was to obtain a qualitative understanding of the differences in the clustering of interstitials between W–Mo and pure W during the post-cascade evolution of defects in these materials.

Lastly, we studied the binding energies of first-nearest-neighbour (1NN) divacancies in pure W and various compositions of W-Mo at 0~K, in lattices that, when devoid of vacancies, consisted of 432 atoms. 
The binding energy of a divacancy was defined to be:
\begin{equation}
    E_{\text{bind, divac}} = E_{\text{form, 1}} + E_{\text{form, 2}} - E_{\text{form, divac}}\nonumber\,
\end{equation}
where $E_{\text{form, 1}}$ and $E_{\text{form, 2}}$ are the formation energies of the two constituent vacancies (obtained from lattices with only one of these vacancies), and $E_{\text{form, divac}}$ is the formation energy of the divacancy. The formation energies for single vacancies are given by:
\begin{equation}
    E_{\text{form, j}} = N_{\text{dist}}\, \left( \frac{E_{\text{dist}}}{N_{\text{dist}}} - \frac{E_{\text{undist}}}{N_{\text{undist}}}\right) \, ,\, j\in \{1, 2\}\, ,
\end{equation}
where $E$ denotes the total potential energy and $N$ the total number of particles of the system specified by the subscripts; the subscript \textit{dist} (disturbed) denotes the system with the vacancy, and \textit{undist} (undisturbed) the defect-free system.

The divacancy formation energy is given by:
\begin{equation}
    E_{\text{form, divac}} = E_{\text{dist}} - E_{\text{undist}} +  2\, \frac{E_{\text{undist}}}{N_{\text{undist}}}\, ,
\end{equation}
where the subscript \textit{dist} now refers to the system with the 1NN divacancy.

For every composition of the W-Mo alloys, we inserted a 1NN divacancy into 15 randomly-generated lattices (30 lattices for the W--Mo-EAM). 
As the binding energy of a 1NN divacancy depends on the chemical composition  of its surroundings, this analysis does not provide a definitive answer to the binding energies of a random W-Mo alloy. 
Rather, the analysis is done to ascertain what effect the addition of Mo to W has on the stability of divacancies.

For comparison, we also computed the divacancy binding energy in DFT for the 50--50 W-Mo composition. Due to computational reasons, we used a smaller lattice (128 atoms) and computed the average of 5 different randomly generated lattices. We used the \textsc{vasp} DFT code~\cite{kresse_ab_1993,kresse_efficient_1996} with projector augmented-wave potentials~\cite{blochl_projector_1994} (\texttt{\_sv} in \textsc{vasp}), the PBE generalized gradient approximation exchange-correlation functional~\cite{perdew_generalized_1996}, 500 eV cutoff energy for the plane-wave basis, 0.15 Å$^{-1}$ maximum $k$-point spacing on Monkhorst-Pack grids~\cite{monkhorst_special_1976}, and 0.1 eV Methfessel-Paxton smearing~\cite{methfessel_high-precision_1989}. These DFT settings are the same as the ones used for generating the training data for GAP and tabGAP~\cite{W-Mo_GAP,tabGAP}.

\subsection{Cluster analysis}
After a cascade, any given two defects in the simulation cell were considered to belong to the same cluster if they were separated by a chosen cut-off distance. The definitions of the cut-off radii for interstitial and vacancy clusters are the same as in Ref.~\cite{cut-off_radii}; for interstitial clusters, the cut-off radius is $\left( r_{\text{3NN}} + r_{\text{4NN}} \right)~\slash~2$, and for vacancy clusters  $\left( r_{\text{2NN}} + r_{\text{3NN}}~\right)~\slash~2$, where
the distance to the $k$th nearest neighbour is $r_{kNN}$. 
The cut-off radii depend on the lattice constant of the cell, which for W-Mo was set to $3.1738~\text{Å}$, as the lattice constants yielded by all three potentials differed from this by less than $1~\%$. 
The lattice constant for equiatomic W-Mo at 300~K as predicted by tabGAP is 3.1800~\aa, GAP 3.179~\aa, and W--Mo-EAM 3.160~\aa. The experimental lattice constant for the 50-50 W-Mo system is roughly 3.16~\aa~\cite{nagender-naiduMoMolybdenumTungstenSystem1984}. The good agreement of the EAM lattice constant with experiment is because of the explicit fitting of the potential to the experimental values, whereas the present GAP-based potentials use the PBE exchange-correlation functional in DFT, which is known to overestimate lattice constants~\cite{haas_calculation_2009}.
For \textit{pure W}, the lattice constant at $300~\text{K}$ given by tabGAP is $3.1892~\text{Å}\,$, W--Mo-EAM $3.1714~\text{Å}$, and AT-ZN EAM 3.1659~Å.



\section{Results and discussion}\label{sec:results}


\subsection{Defect formation and mobility}\label{sec:defects}

\begin{figure}[h!]
    \centering
    \includegraphics[width=.85\columnwidth]{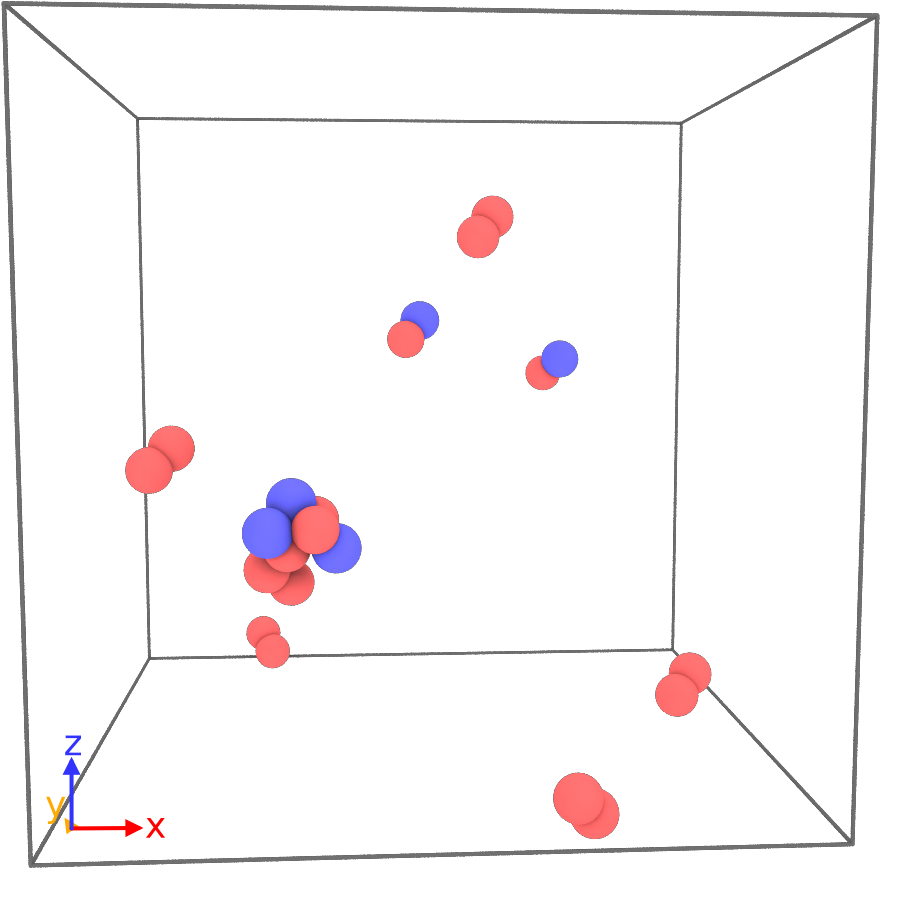}
    \caption{An exemplary snapshot of the simulation cell with interstitials produced in a 10-keV (tabGAP) cascade simulation in W-Mo. Single split-interstitials are aligned with different $\left\langle 111 \right\rangle$ directions as expected in a BCC lattice. In the interstitial cluster (downleft from the center of the box), all the interstitials are aligned in one of the $\left\langle 111 \right\rangle$ directions (in the snapshot, it is $\left[ \bar{1} \bar{1} 1 \right]$). Here the blue atoms are W, and the red atoms are Mo. The box borders are downscaled from the original size borders of the simulation cell to enclose the region with the generated interstitials only. The x, y and z axes are aligned with the $\left[ 1 0 0 \right]$, $\left[ 0 1 0 \right]$, and $\left[ 0 0 1 \right]$ crystallographic directions, respectively.}
    \label{fig:interstitial-render}
\end{figure}

The interstitials produced in a 10-keV (tabGAP) cascade simulation in W-Mo are shown in Fig.~\ref{fig:interstitial-render}. One can see that single split-interstitials are oriented along different $\left\langle 111 \right\rangle$ directions, while in the SIA cluster (centre of the snapshot), the interstitials are aligned along $\left[ \bar{1} \bar{1} 1 \right]$ direction parallel to one another, which is consistent with the shape of the clusters observed earlier in tungsten~\cite{Ale16}. 
%

\begin{figure}[h!]
    \begin{subfigure}{\columnwidth}
        \includegraphics[width=\linewidth]{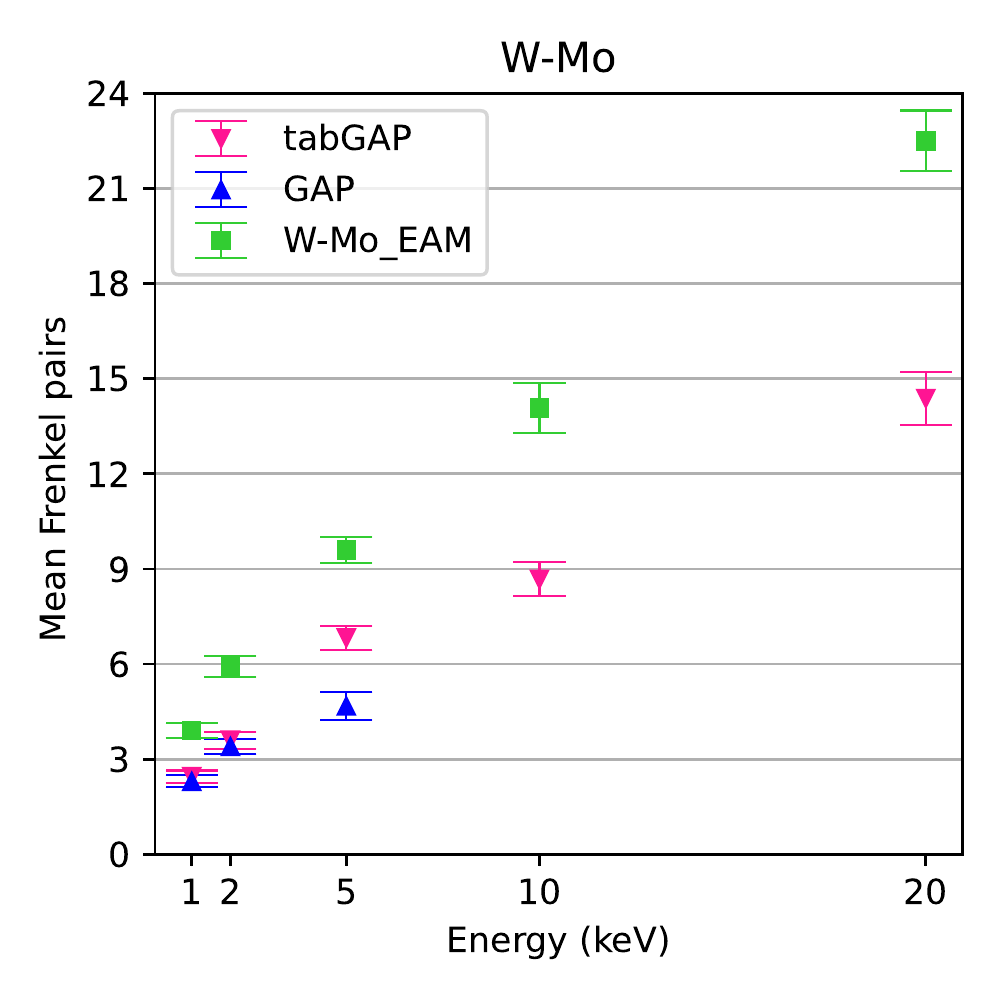}
        \caption{}
        \label{fig:W-Mo_ndefects}
    \end{subfigure}
    
    \begin{subfigure}{\columnwidth}
        \includegraphics[width=\linewidth]{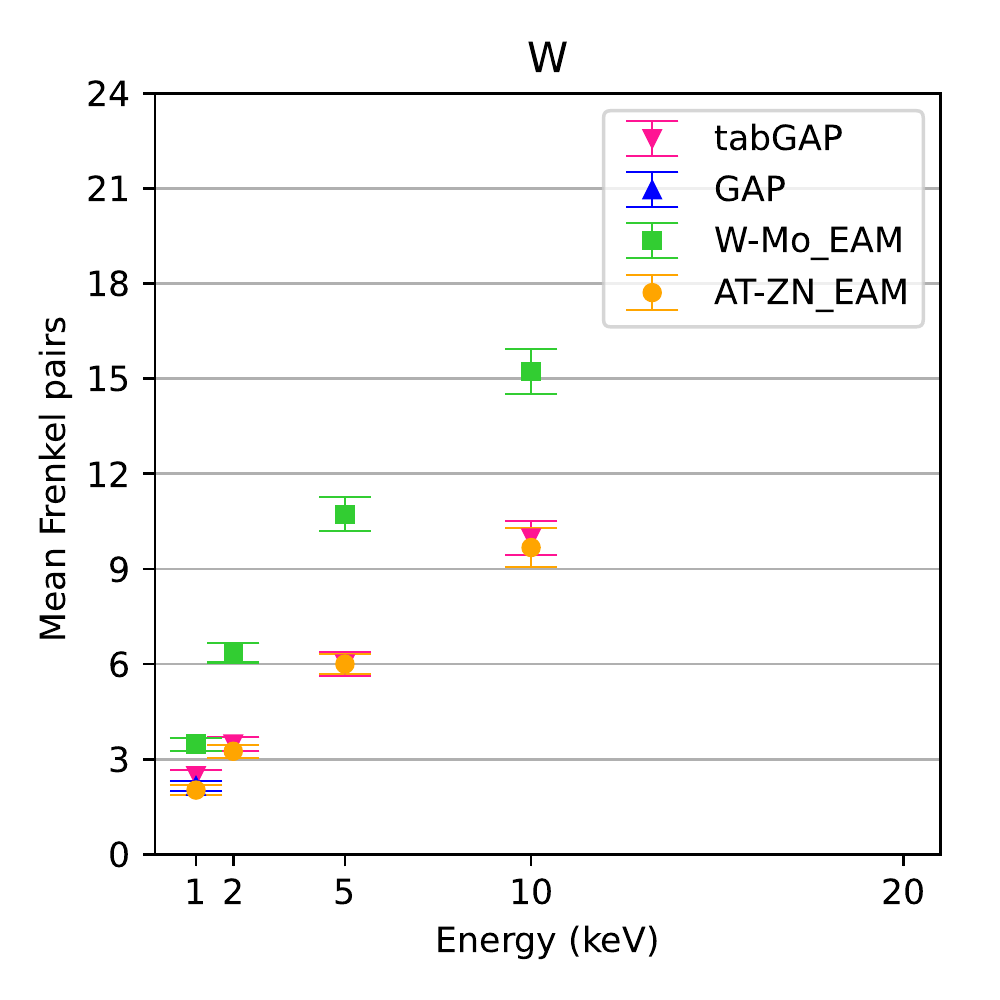}
        \caption{}
        \label{fig:W_ndefects}
    \end{subfigure}
    \caption{Mean number of Frenkel pairs with respect to PKA energy for cascades in (a) W-Mo and (b) pure W. The vertical bars indicate the standard error.
    Results from all simulations, even those that ended with zero defects, were included in the averages and the errors thereof.}
    \label{fig:n_defects}
\end{figure}

\begin{figure*}[h!]
    \begin{subfigure}{.7\columnwidth}
        \includegraphics[width=\columnwidth]{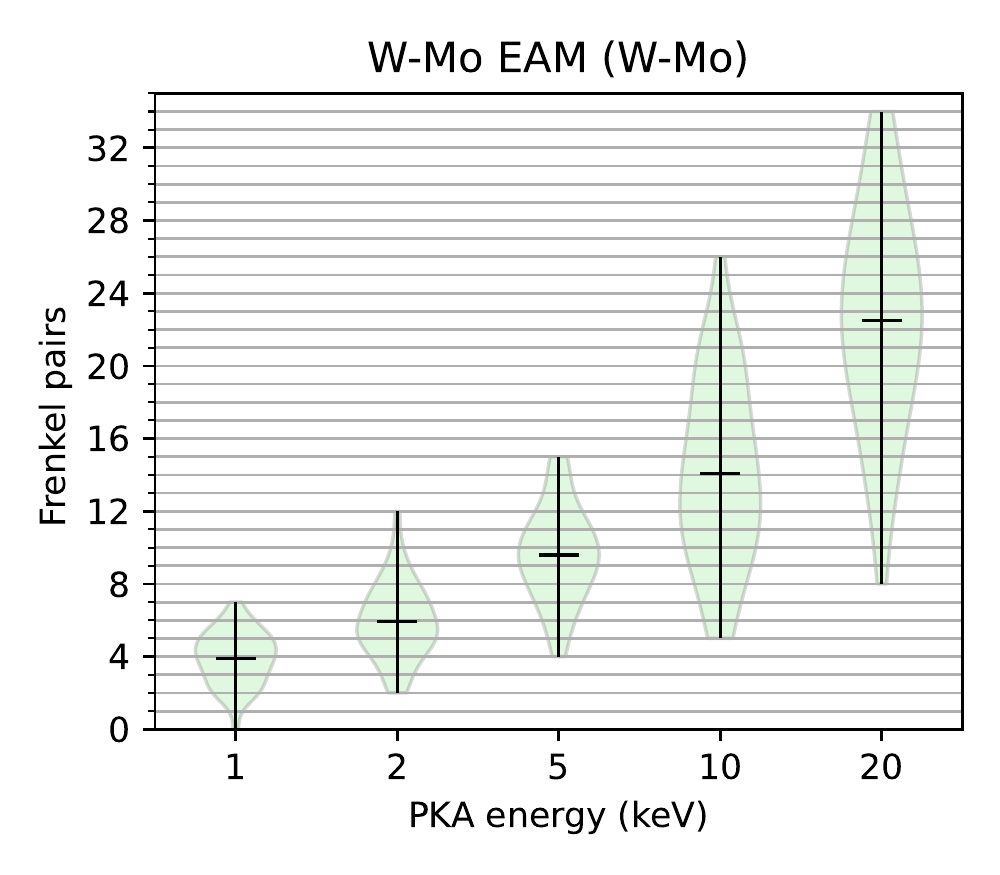}
        \caption{}
        \label{fig:EAM_violin_W-Mo}
    \end{subfigure}
    \begin{subfigure}{.7\columnwidth}
        \includegraphics[width=\columnwidth]{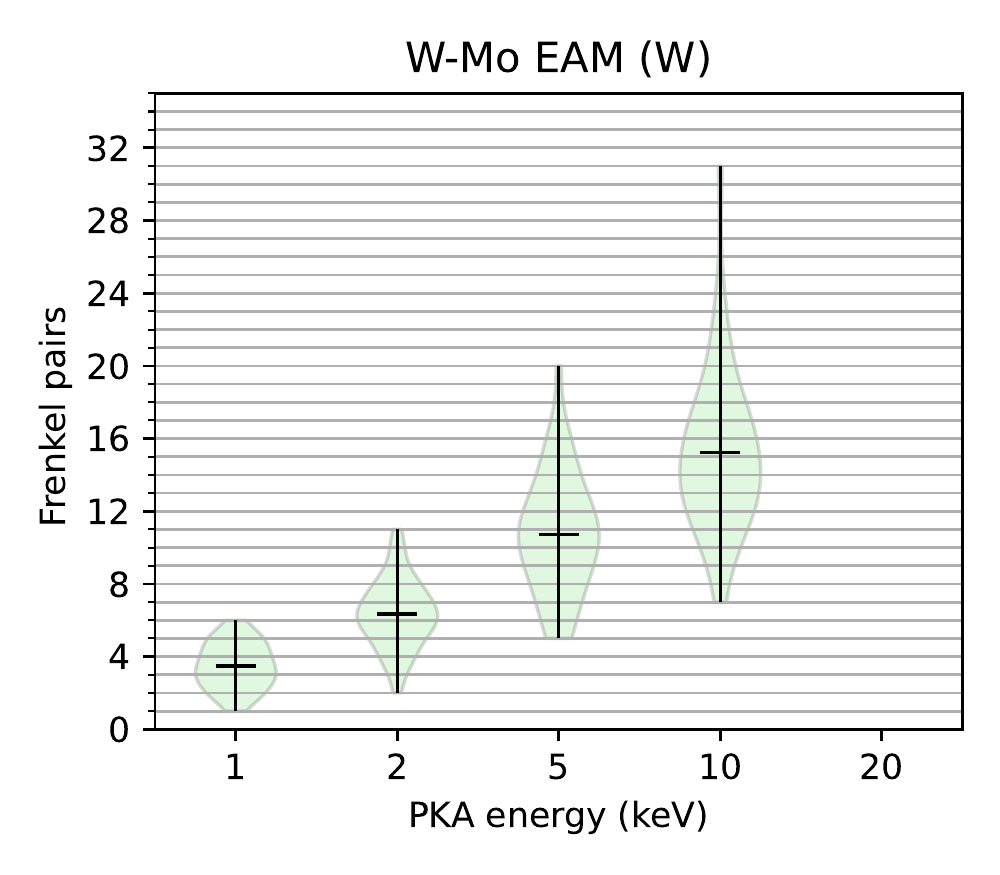}
        \caption{}
        \label{fig:EAM_violin_W}
    \end{subfigure}
    \\
    \begin{subfigure}{.7\columnwidth}
        \includegraphics[width=\columnwidth]{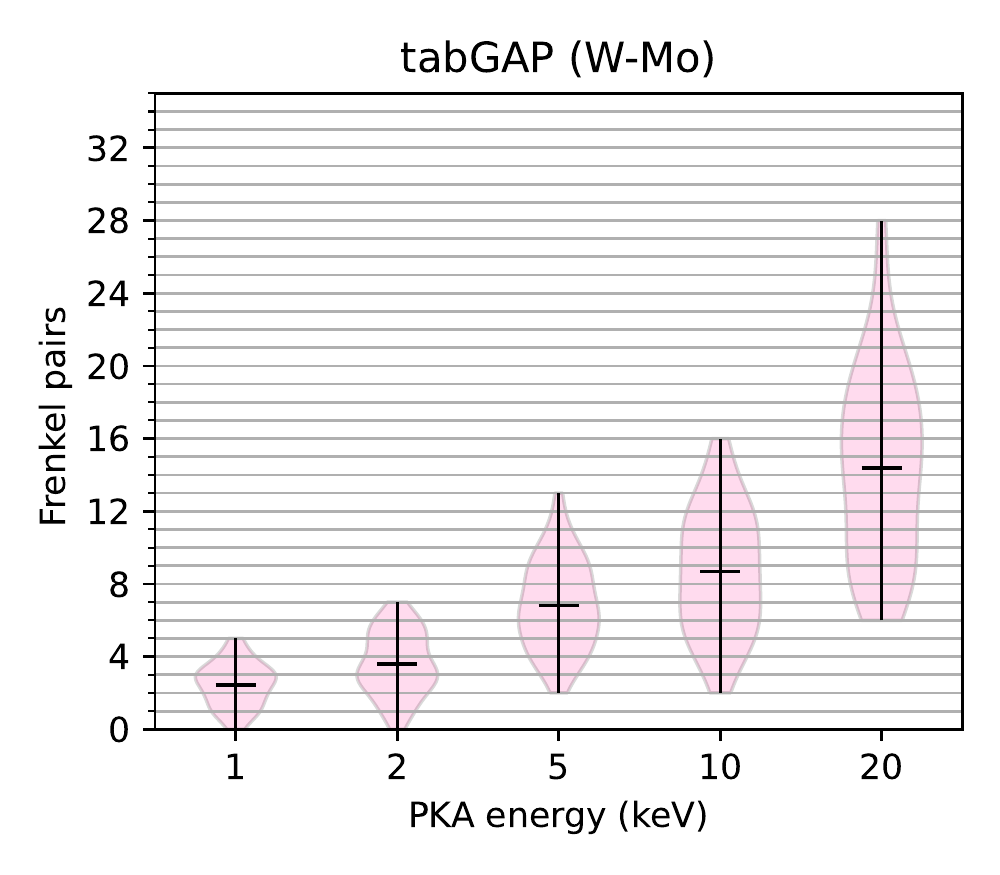}
        \caption{}
        \label{fig:tabGAP_violin_W-Mo}
    \end{subfigure}
    \begin{subfigure}{.7\columnwidth}
        \includegraphics[width=\columnwidth]{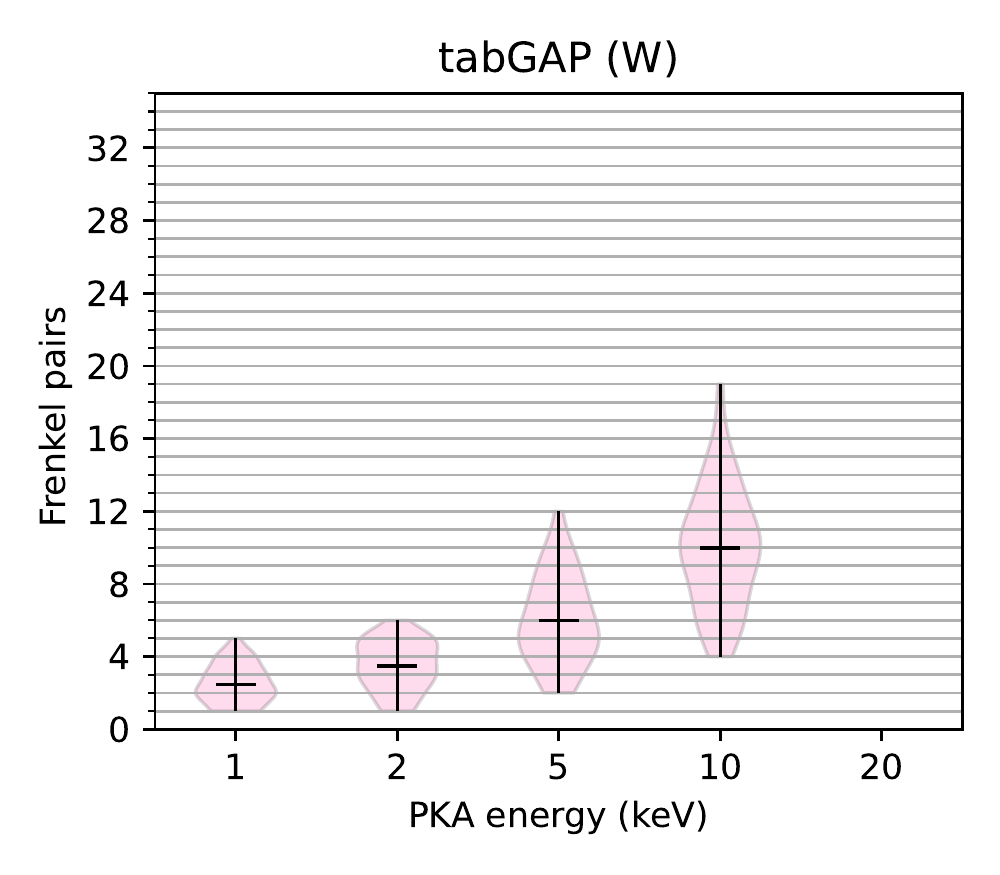}
        \caption{}
        \label{fig:tabGAP_violin_W}
    \end{subfigure}
    \\
    \begin{subfigure}{.7\columnwidth}
        \includegraphics[width=\columnwidth]{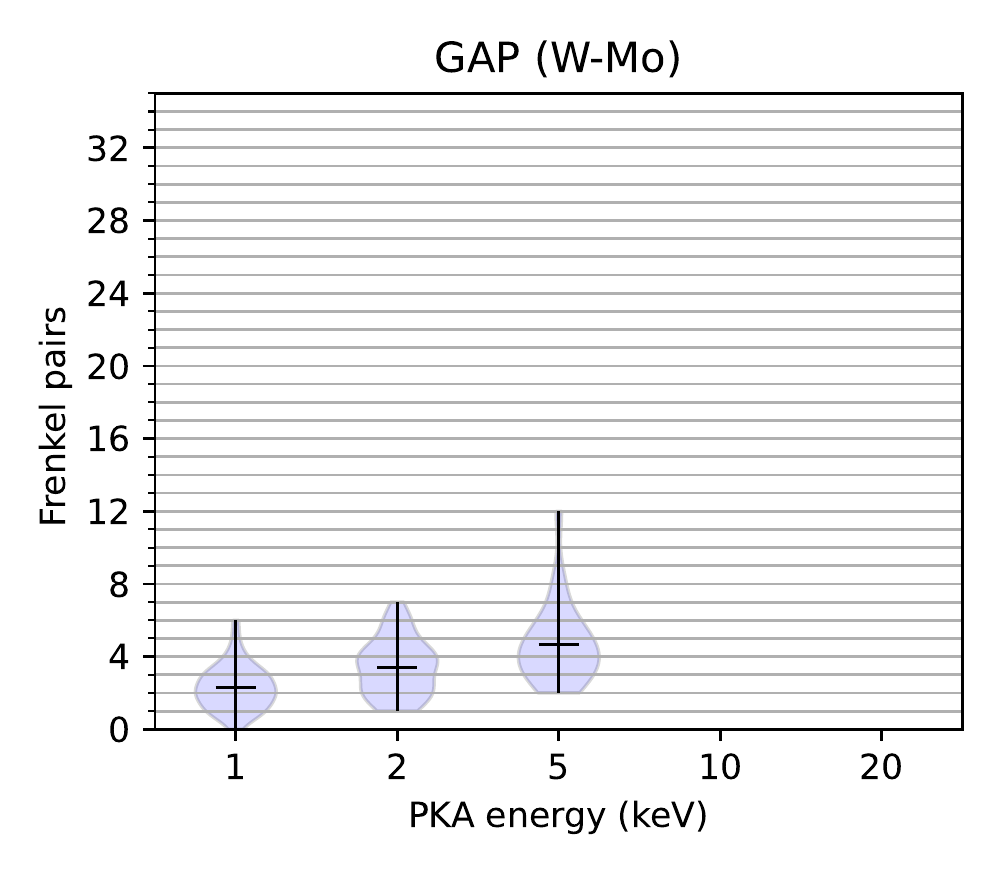}
        \caption{}
        \label{fig:GAP_violin_W-Mo}
    \end{subfigure}
        \begin{subfigure}{.7\columnwidth}
        \includegraphics[width=\columnwidth]{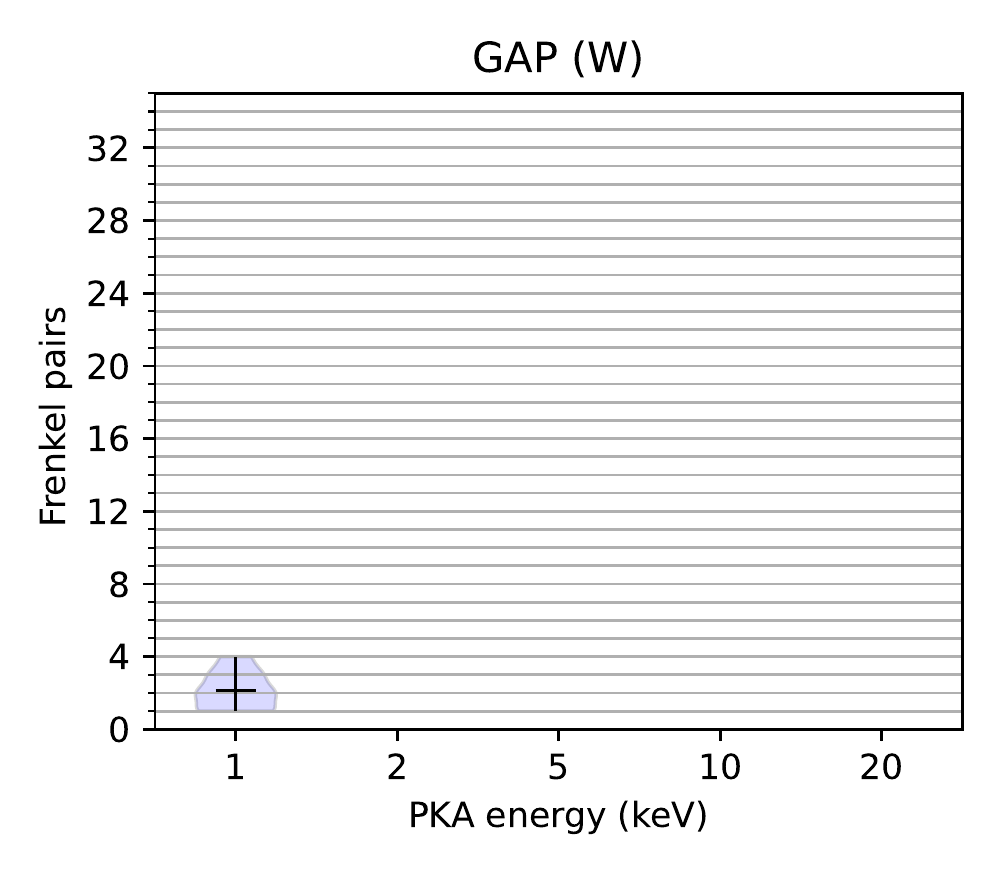}
        \caption{}
        \label{fig:GAP_violin_W}
    \end{subfigure}
    \\
    \begin{subfigure}{.7\columnwidth}
        \includegraphics[width=\columnwidth]{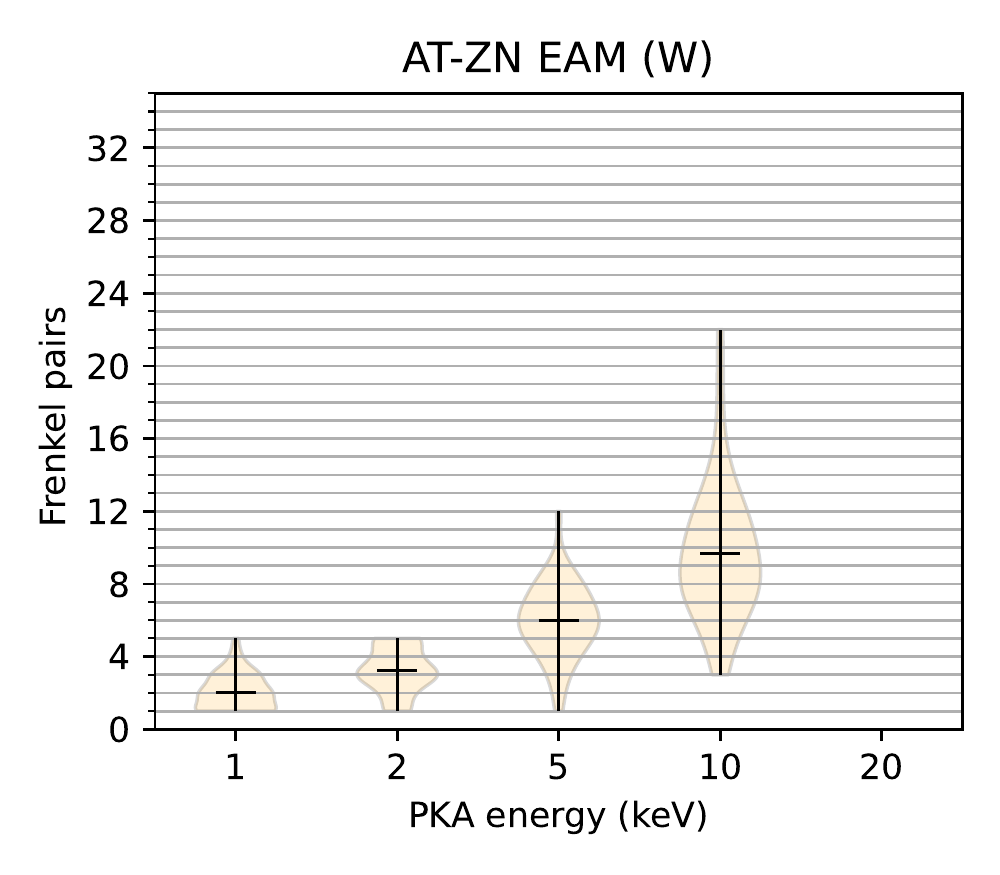}
        \caption{}
        \label{fig:AT_ZN_violin_W}
    \end{subfigure}
    \caption{
    Frenkel pair (FP) violin plots. The violin shapes show the distributions of probability density to create a corresponding number of FPs (y-axis) at a given energy of the recoil (x-axis). W–Mo EAM denotes the EAM developed for W–Mo. The horizontal grid guides the eye to correlate  the possible values of the number of FPs with the violin. The vertical line inside the violins points to the recoil energy on the x-axis for which the violin graph was generated. The horizontal lines inside the violins are the means of the probability density distributions.
    }
    \label{fig:violinplots}
\end{figure*}

The mean number of Frenkel pairs as a function of the PKA energy is presented in  Fig.~\ref{fig:n_defects} for both materials. 
It should be noted that the results of all simulations were included when evaluating averages and standard errors related to the number of defects, even those that ended with no defects.
Information on how the results in individual simulations are distributed around the mean is illustrated in Fig.~\ref{fig:violinplots}.

As shown in Fig.~\ref{fig:n_defects}, tabGAP and GAP produce a comparable number of defects. At 5~keV in W-Mo, however, tabGAP produces slightly more defects, though, given the standard error, the difference can be as low as about 1 to 2 defects. 
The W--Mo-EAM, on the other hand, produces significantly more defects across the board, in both W-Mo and W. 
This is likely due to the threshold displacement energies reported in Ref.~\cite{CHEN2020152020} being too low for the present W--Mo-EAM, although results were only reported for pure Mo.
We also observe that the predictions made by the AT-ZN EAM and tabGAP for the mean number of surviving defects are similar, although the numbers predicted by tabGAP are slightly higher.
\\
\\
An interesting property of W-Mo manifests itself in the violin plots (Figs.~\ref{fig:EAM_violin_W-Mo}, \ref{fig:tabGAP_violin_W-Mo}, and \ref{fig:GAP_violin_W-Mo}), 
namely exhibiting recombination of all defects to some extent at lower PKA energies; even in one W--Mo-EAM 1-keV simulation,  the cell completely recovered from the damage after the cascade had cooled down. In W, defect recombination was not observed in any of the tested PKA energies, though looking at Fig.~\ref{fig:n_defects}, tabGAP and GAP describe W as producing roughly the same number of defects as W-Mo (given the standard errors), whereas W--Mo-EAM predicts a greater mean number of defects in W than W-Mo.

\begin{figure}[h!]
    \begin{subfigure}{\columnwidth}
      \includegraphics[width=\columnwidth]{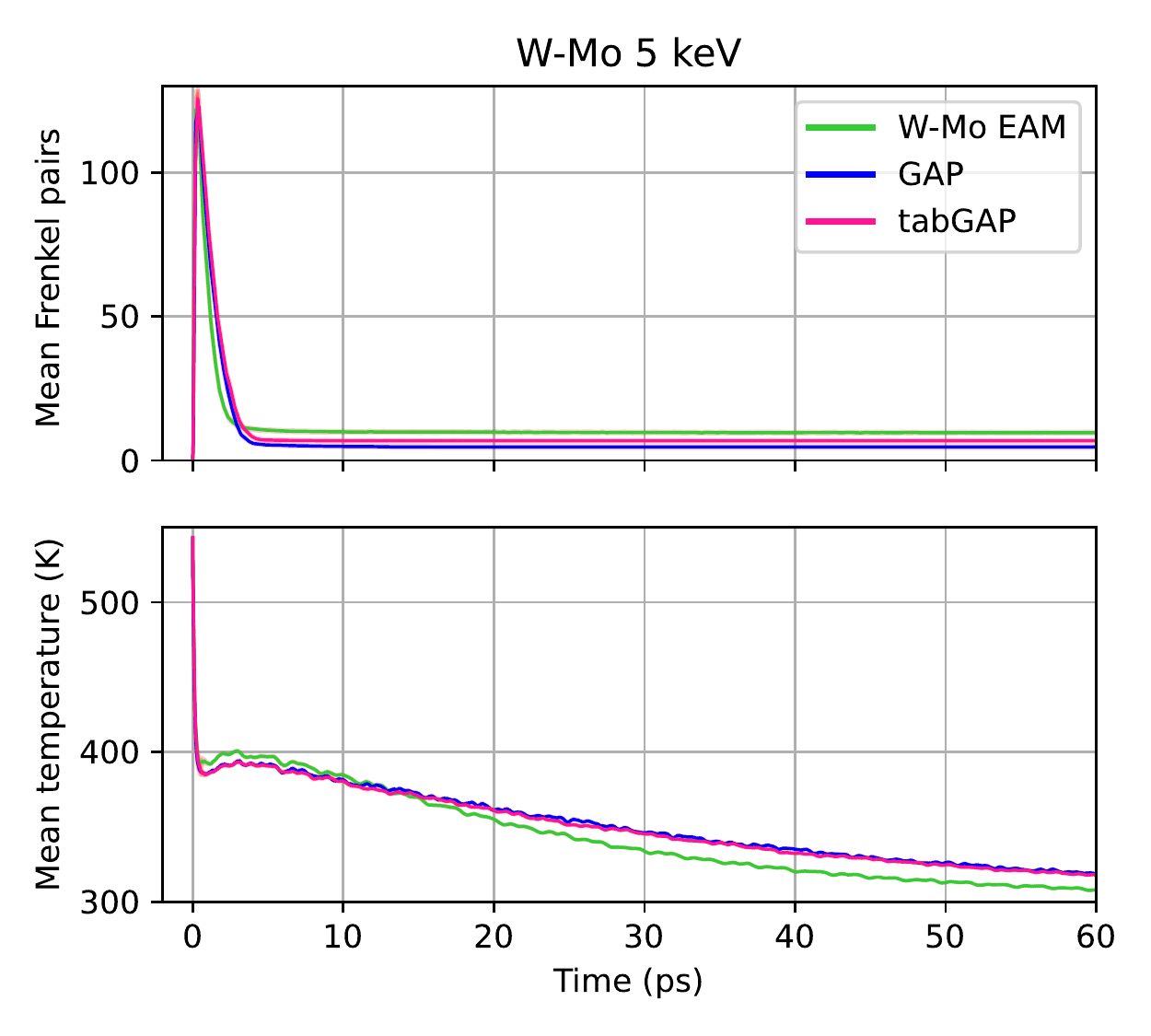}        
      \caption{}
    \end{subfigure}\\
    \begin{subfigure}{\columnwidth}
     \includegraphics[width=\columnwidth]{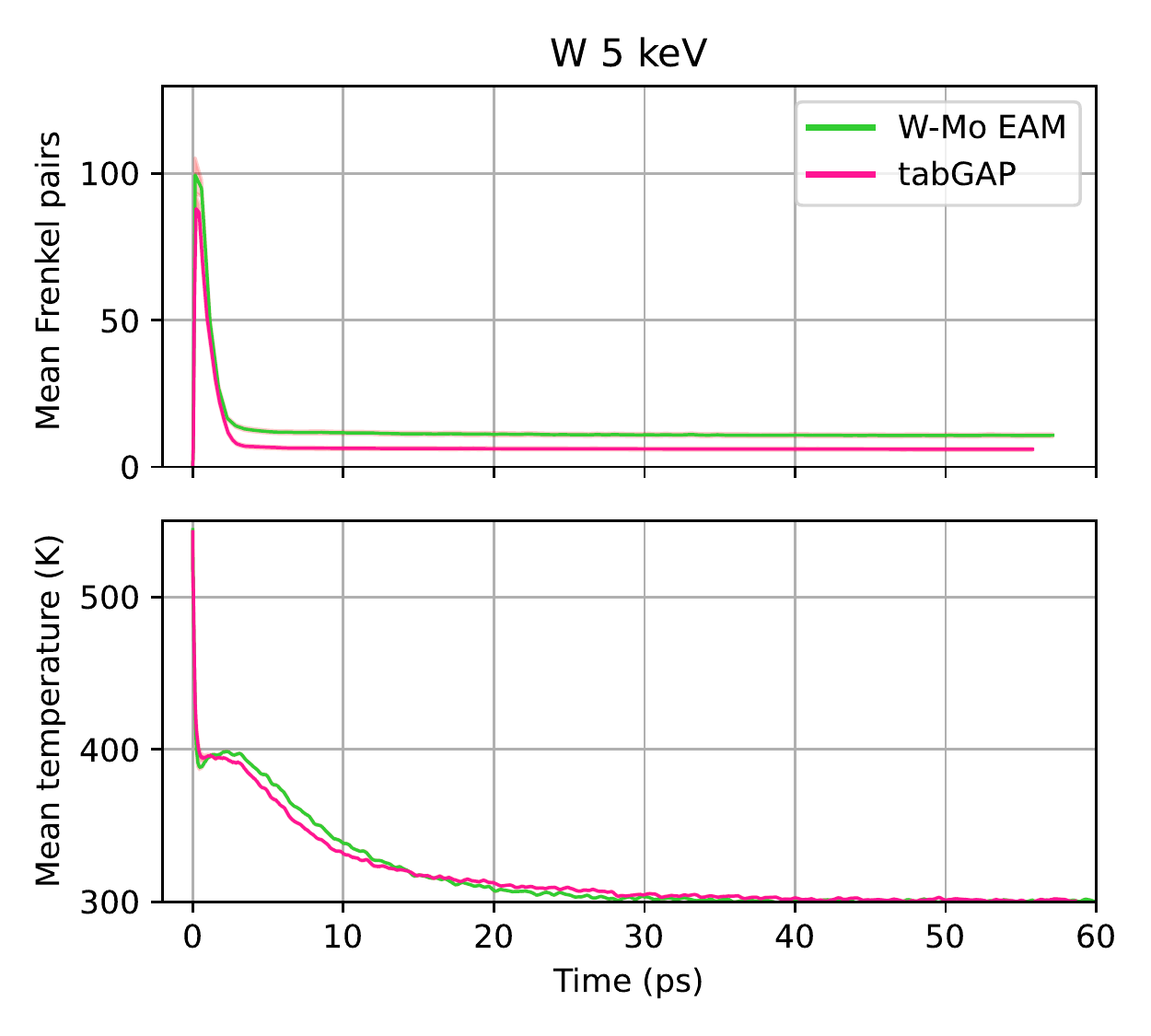}
     \caption{}
    \end{subfigure}

    \caption{Mean defect-formation and temperature plots for W-Mo and W 5-keV simulations. 
    The top plots show the mean number of Frenkel pairs, and the bottom plots show the mean temperature, both with respect to time. Standard error, albeit very small, is represented by a shaded red area. The x-axis (time) is shared among the defect and temperature plots. 
    The x-axis has been limited to 60~ps for clarity.
    Results from all simulations, even those that ended with zero defects, were included in the averages and the errors thereof.}
    \label{fig:dtt}
\end{figure}

In Fig.~\ref{fig:dtt}, one can discern the temporal evolution of temperature and defect formation in 5-keV W-Mo and W cascade simulations. We note that the temperature during the highly non-equilibrium peak of the cascade is not a conventional equilibrium temperature, but a measure of the average kinetic energy $E_ \mathrm{kin}$ of the system transformed to temperature $T$ using $E_\mathrm{kin}=\frac{3}{2} Nk_{\text{B}} T$. The absolute value of the temperature is not meaningful, as it depends on the number of atoms $N$ in the simulation cell. However, the time dependence of $T$ is a good illustration of the duration of the non-equilibrium phase of a collision cascade. 

On the account of Fig.~\ref{fig:dtt}, it is apparent that defects stop being produced shortly after the initial spike in temperature, caused by the development of the cascade.
W-Mo demonstrates a more efficient recombination of defects produced during the initial phase of the cascades than W; W-Mo has an initial spike of around 130 defects, whereas W has around 100 defects, yet both materials end up with roughly the same mean number of defects.
Furthermore, the temperature is removed from the W-Mo cell more efficiently by the W--Mo-EAM potential compared to GAP and tabGAP, both of which had similar predictions.
This is apparent from the comparison of the temperature evolution in the simulation cell after the active cascade phase under the same boundary conditions in all three potentials.
This discrepancy may be explained not only by different lattice thermal conductivities but also by cascade size and shape.

The analysis of the interstitial-mobility simulations revealed that interstitials at a given temperature in W-Mo are far less mobile than in W, where interstitials had effectively no movement even at $600~\text{K}$. At a temperature of $1200~\text{K}$, the mobility W-Mo interstitials rivalled the mobility pure-W interstitials had at $600~\text{K}$. The interstitials were observed to migrate mainly in a crowdion $\left\langle 111 \right\rangle$ direction in both W-Mo and W.

Considering that interstitials in W-Mo at $600~\text{K}$ are practically immobile on the MD time scale, and that the temperature even at 5~keV drops far below 600~K during the first few picoseconds, the shorter run-time of GAP 5~keV (shortest was 71~ps) most likely had no effect on defect formation and clustering. In pure W, the temperature was observed to decrease faster than in W-Mo, having reached 300~K long before 60~ps had transpired in 5-keV simulations, as indicated in Fig.~\ref{fig:dtt}. This indicates that the lattice thermal conductivity is significantly higher in pure W than in random W-Mo alloys.

\subsection{Defect clustering}
\begin{figure}[h!]
    \centering
    \includegraphics[width=\columnwidth]{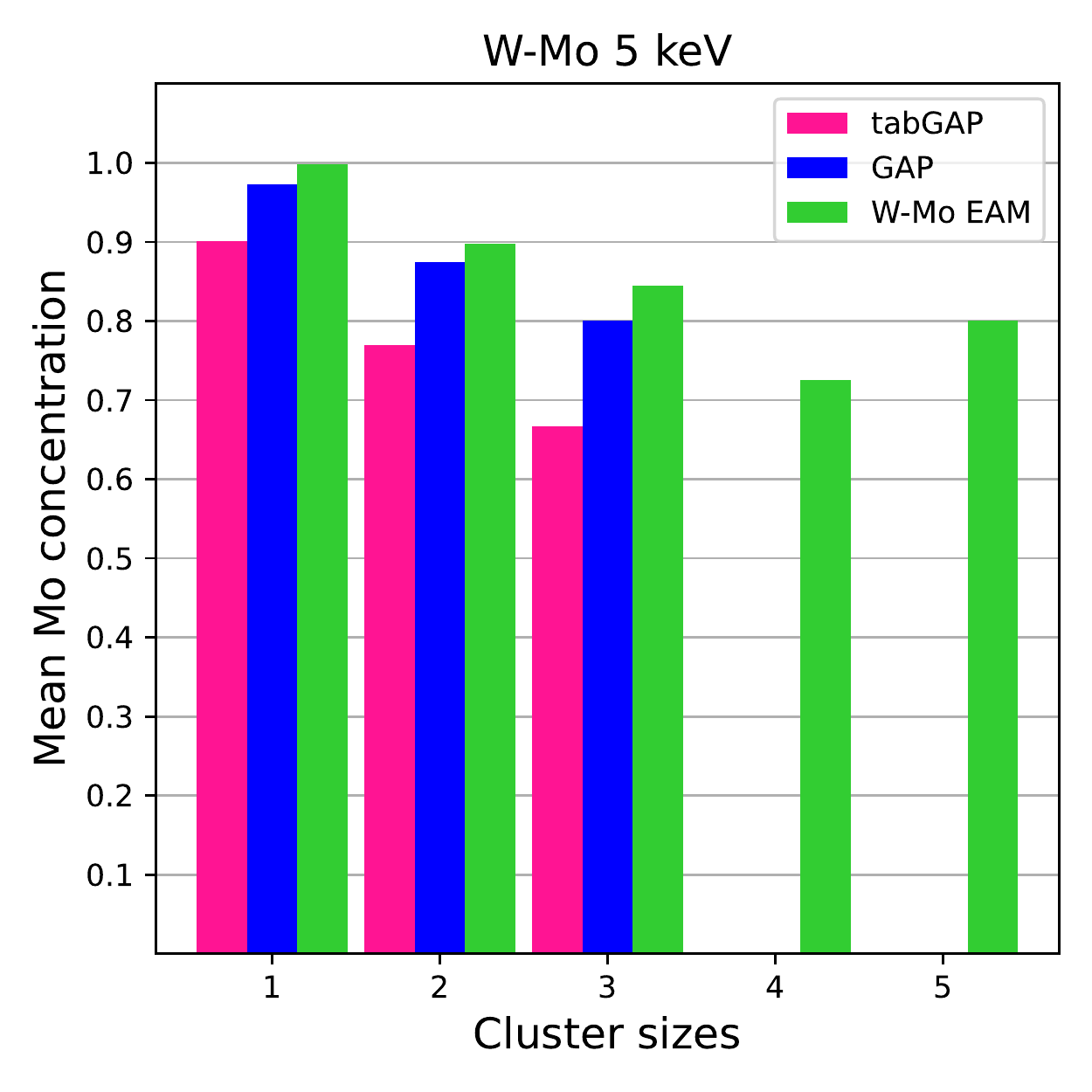}
    \caption{Average fraction of Mo found in interstitial clusters in 5~keV simulations, with respect to cluster size. W-Mo EAM is the EAM developed for W-Mo.
    Size-1 clusters are single split-interstitials, comprising two atoms.}
    \label{fig:Mo-frac_5_keV}
\end{figure}

The Mo concentrations in interstitial clusters of 5-keV simulations are depicted in Fig.~\ref{fig:Mo-frac_5_keV}, wherein Mo-Mo is shown to be the predominant type of split-interstitial.
Moreover, tabGAP clusters have a slightly larger fraction of W than GAP. 

We note that the present tabGAP was trained for Mo-Nb-Ta-V-W, which means a smaller fraction of its training data describes W-Mo interactions than the GAP, which was trained directly for W-Mo alloys. 
Nevertheless, all three potentials agree that Mo atoms are predominant in interstitial clusters in W-Mo.
\\
\\
Statistical distributions of vacancy and interstitial clusters are shown in Fig.~\ref{fig:clusters}. 
More distributions for the remaining tested energies are given in the Supplementary material. 
Given the standard errors, the comparison between the different potentials is satisfactory. Some of the clusters are seen in some potentials, but not in others. Overall, the GAP predicts smaller cluster sizes than the W--Mo-EAM potential and tabGAP.

Fig.~\ref{fig:clusters} shows that in pure W, interstitial clusters are more prevalent than in W-Mo, which is reasonable given the increased mobility that interstitials in W have over those in W-Mo. Differences between W-Mo and W in the clustering of interstitials at PKA energies lower than 5~keV are less consistent. This is due to the overall low probability of the formation of large clusters at these energies, which makes the data noisier and less statistically reliable.

The interstitial clustering in W is similar in both tabGAP and AT-ZN EAM, taking into account the margins of error.
However, W--Mo-EAM predicts that the vacancies cluster more in pure W than in W-Mo, whereas tabGAP predicts the opposite. 
Moreover, the AT-ZN EAM predicts a higher number of vacancy clusters (size~$>$~1) than tabGAP.

\begin{figure}
    \centering
    \includegraphics[width=\columnwidth]{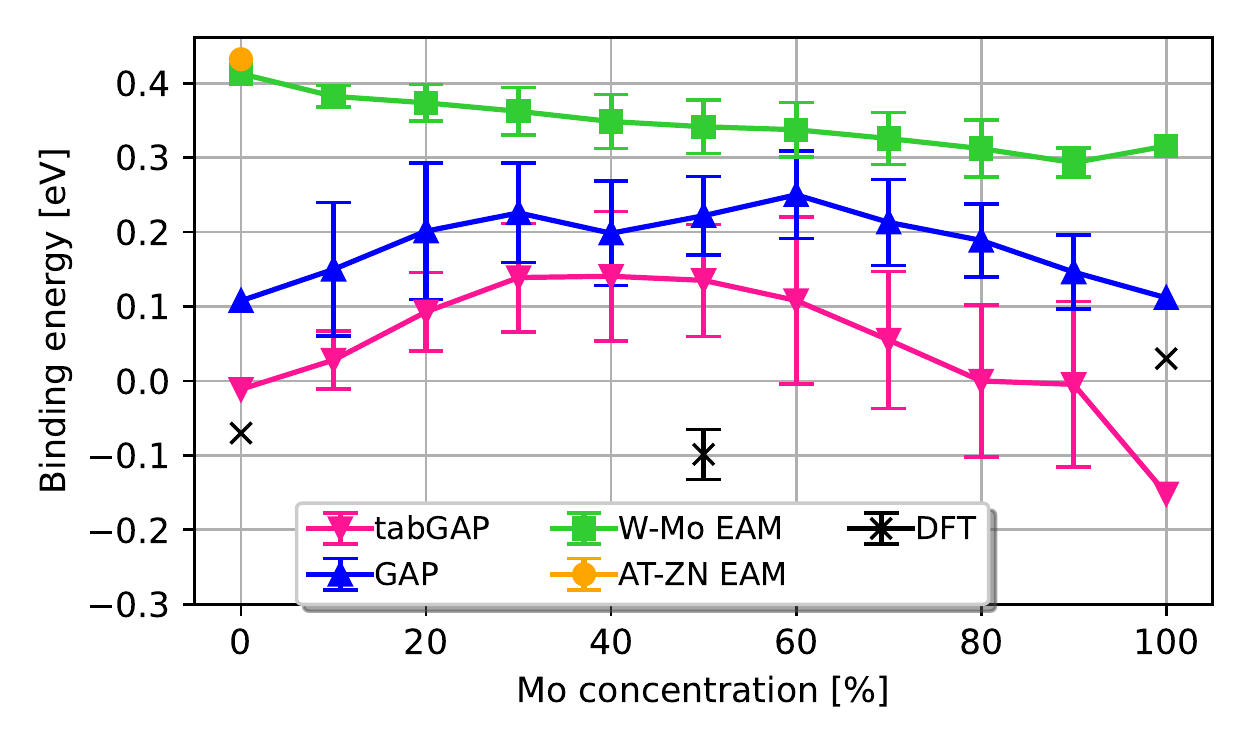}
    \caption{Mean binding energies of 1NN divacancies with respect to the Mo concentration in W-Mo. A negative value indicates that the divacancy is unstable. Aside from the mono-elemental values, which were obtained from a single simulation, each tabGAP/GAP point is the average of 15 different, randomly-generated alloys, and each W--Mo-EAM point is the mean of 30. The vertical lines denote the sample standard error of the mean. AT-ZN EAM has only one measurement, as it is purely a W potential. The DFT values for the monoelemental cases are from Ref.~\cite{byggmastar_gaussian_2020}, whereas the 50-\% point is computed in this work and is the mean of five configurations.}
    \label{fig:divac_bind}
\end{figure}

We note that tabGAP predicts more efficient clustering of vacancies in W-Mo compared to W, which is in agreement with the divacancy binding energy in Fig.~\ref{fig:divac_bind}. However, DFT predicts that divacancies in W-Mo alloys are roughly as unstable as in pure W, suggesting that alloying may not affect vacancy clustering. 
Fig.~\ref{fig:divac_bind} shows that none of the potentials (not even GAP) reproduce the DFT trend for divacancy stability, although, the divacancy binding energies predicted by GAP and tabGAP are much closer to DFT than those of the AT-ZN EAM (for pure W) and the W--Mo-EAM. 
We note that the small magnitudes ($\approx0.1~\text{eV}$) of the binding energies (including the negative binding energies) are much smaller than the kinetic energies in the collision cascades ($>100~\mathrm{eV}$), and hence, they are not expected to have a strong effect on the results of the present study.  
Moreover, it has been shown that, despite their negative binding energy, divacancies are fairly stable in W because of high dissociation energies ($\approx 1.7$ eV)~\cite{heinolaStabilityMobilityDivacancies2017}. For that reason, even in the long-term evolution of defects, this inaccuracy in the binding energy is not expected to affect remarkedly vacancy clustering in these materials, since the energy barriers for vacancy migration are usually over 1.5 eV~\cite{heinolaStabilityMobilityDivacancies2017}. However, for higher accuracy of the description of cluster dynamics in cascades, we recommend to re-train the tabGAP and specifically include the defects of interest to ensure that the machine-learning algorithm sees the corresponding configurations during training.


\begin{figure*}
    \begin{subfigure}{\columnwidth}
        \includegraphics[width=\linewidth]{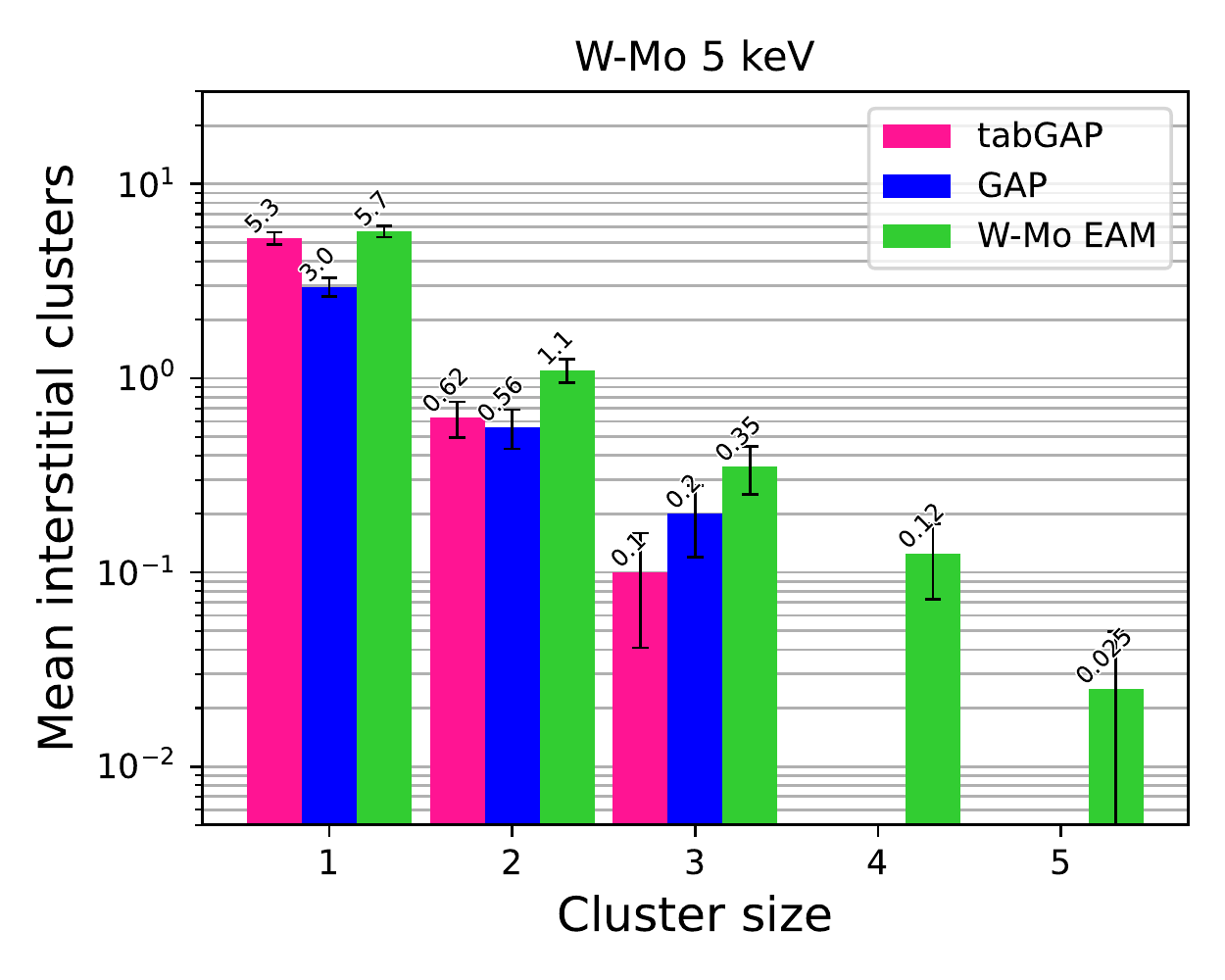}
        \caption{}
        \label{fig:W-Mo_intcluster_5_keV}
    \end{subfigure}
    \begin{subfigure}{\columnwidth}
        \includegraphics[width=\linewidth]{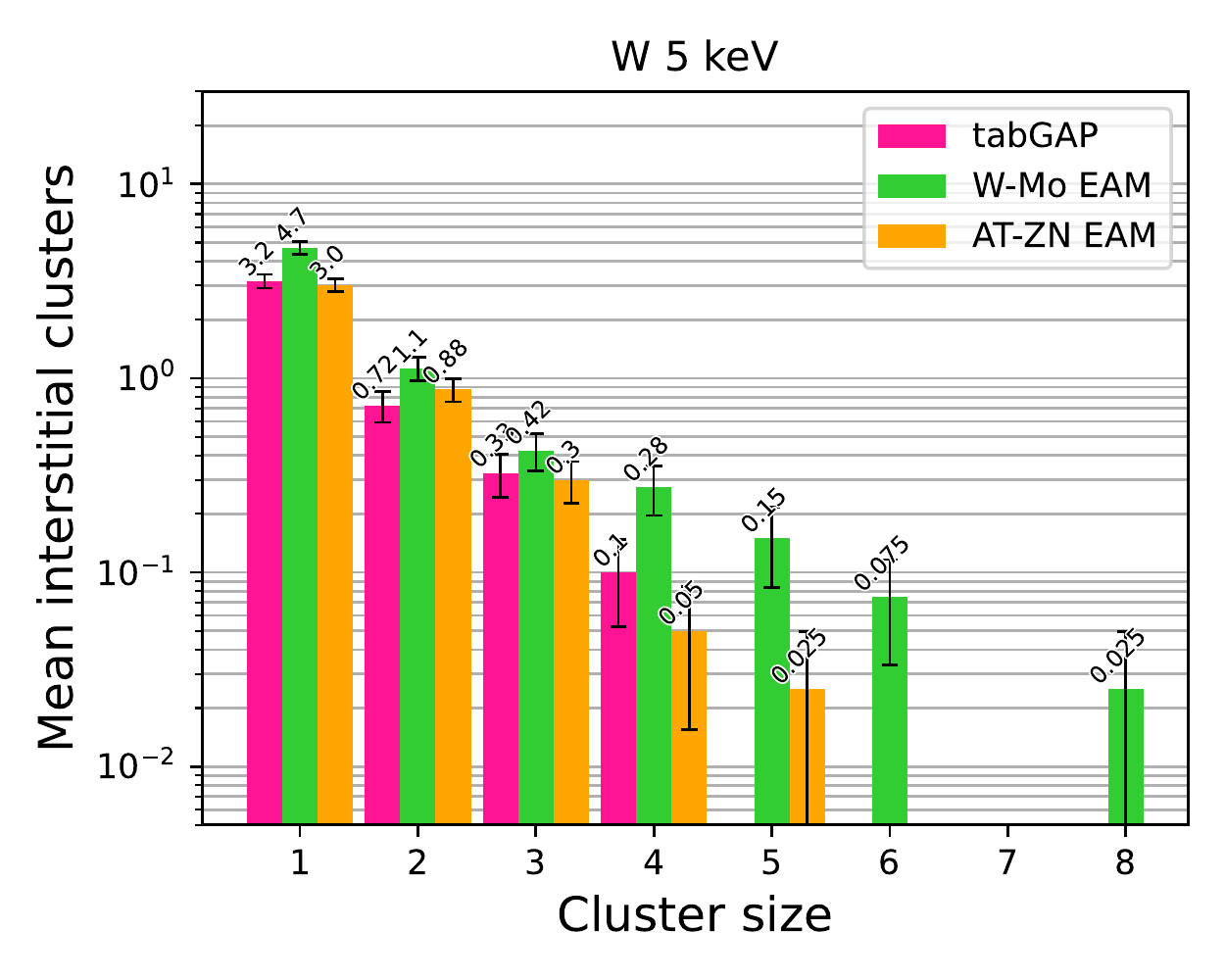}
        \caption{}
        \label{fig:W_intcluster_5_keV}
    \end{subfigure}
    \\
    \begin{subfigure}{\columnwidth}
        \includegraphics[width=\linewidth]{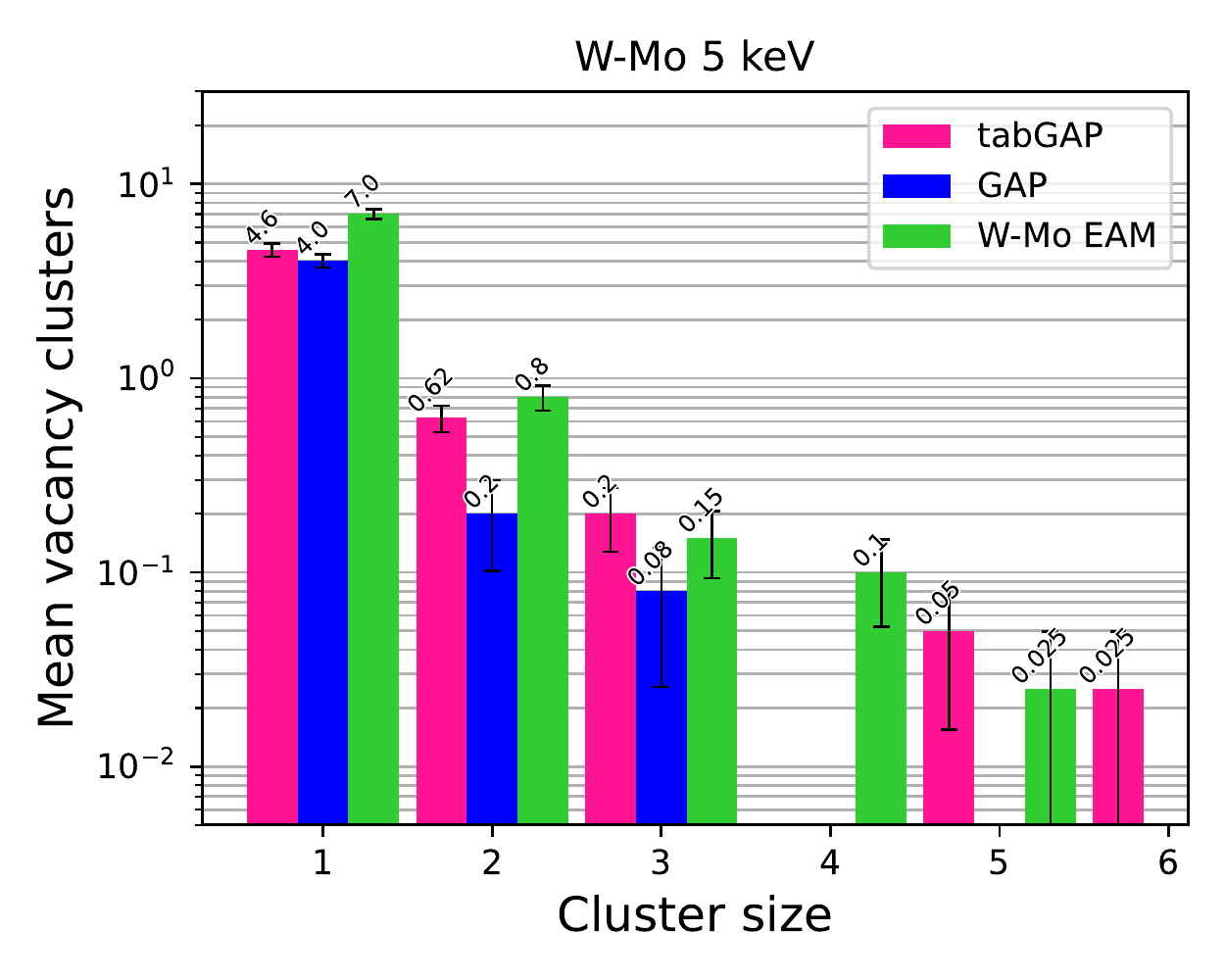}
        \caption{}
        \label{fig:W-Mo_vaccluster_5_keV}
    \end{subfigure}
    \begin{subfigure}{\columnwidth}
        \includegraphics[width=\linewidth]{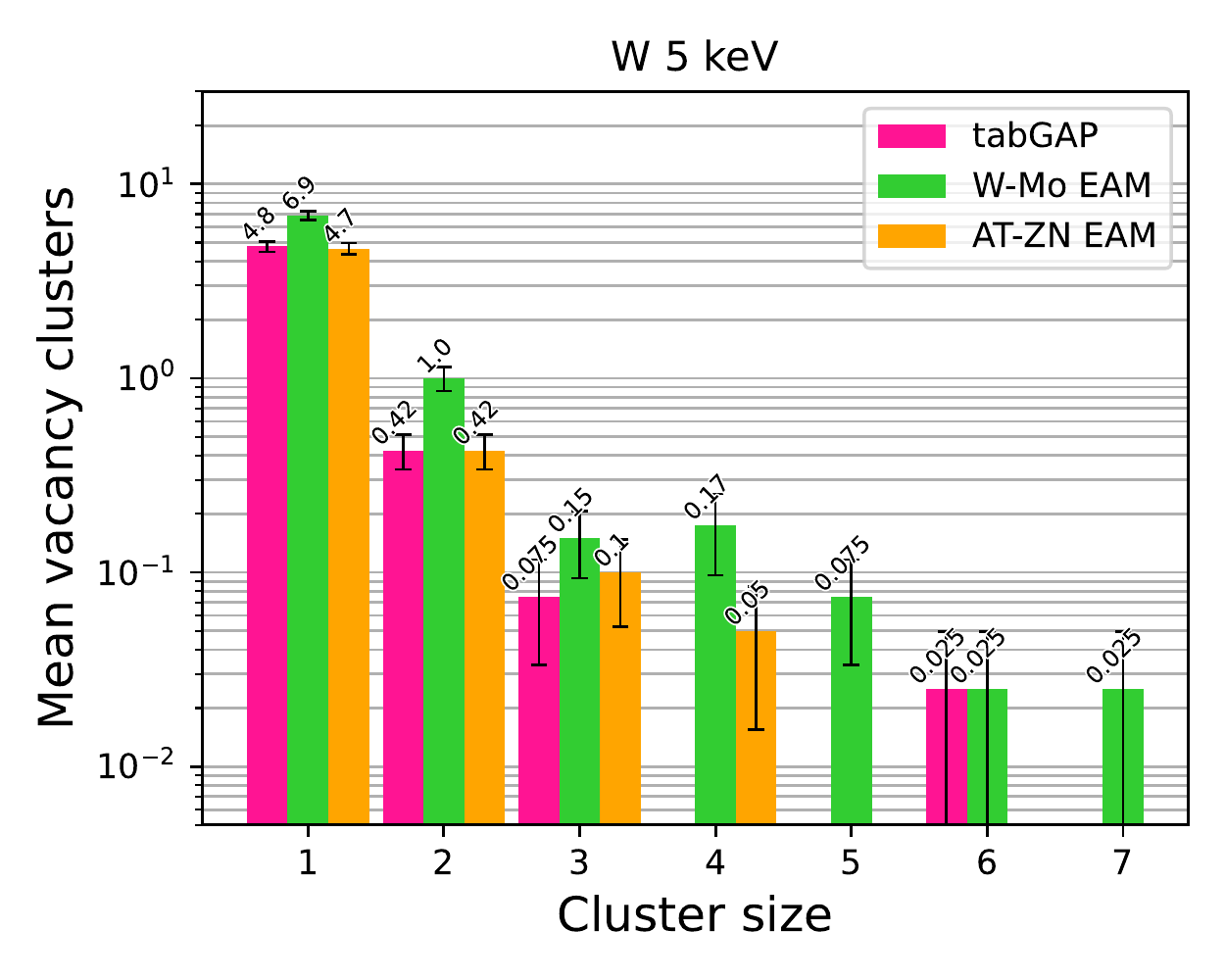}
        \caption{}
        \label{fig:W_vaccluster_5_keV}
    \end{subfigure}
    \\
    \begin{subfigure}{\columnwidth}
        \includegraphics[width=\linewidth]{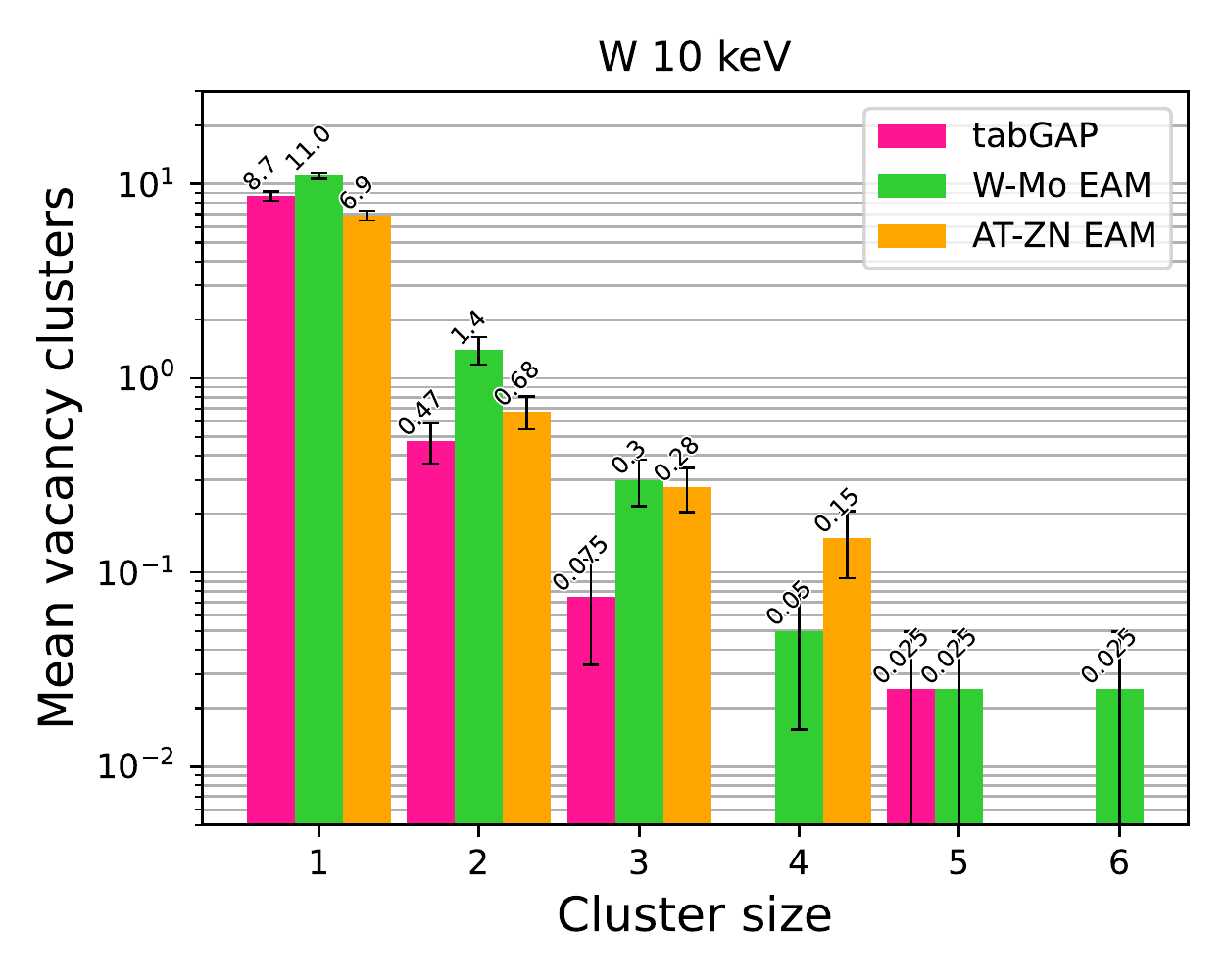}
        \caption{}
        \label{fig:W_vaccluster_10_keV}
    \end{subfigure}
    \begin{subfigure}{\columnwidth}
        \includegraphics[width=\linewidth]{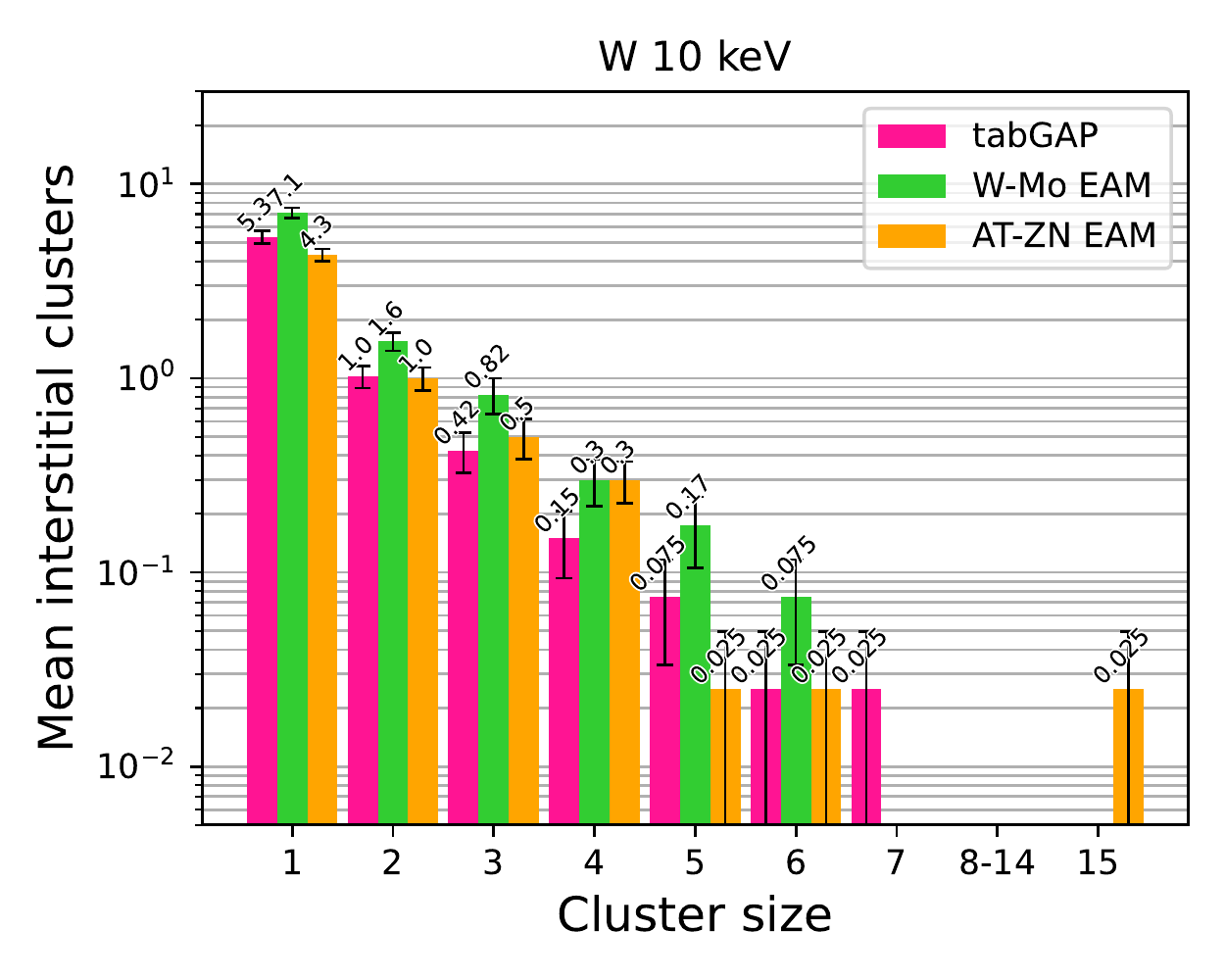}
        \caption{}
        \label{fig:W_intcluster_10_keV}
    \end{subfigure}

    \caption{Histograms of defect cluster size distributions. W--Mo EAM refers to the EAM made for W-Mo. The y-axis is the number of clusters, the x-axis is the cluster size. The numbers atop the bars express their y-values and have been included for clarity due to the usage of a logarithmically scaled y-axis. The vertical line at each bar gives the corresponding standard error (standard errors lower on the y-axis appear significantly larger due to the logarithmic y-axis).}
    \label{fig:clusters}
\end{figure*}

\begin{figure*}
    \begin{subfigure}{\columnwidth}
        \includegraphics[width=\linewidth]{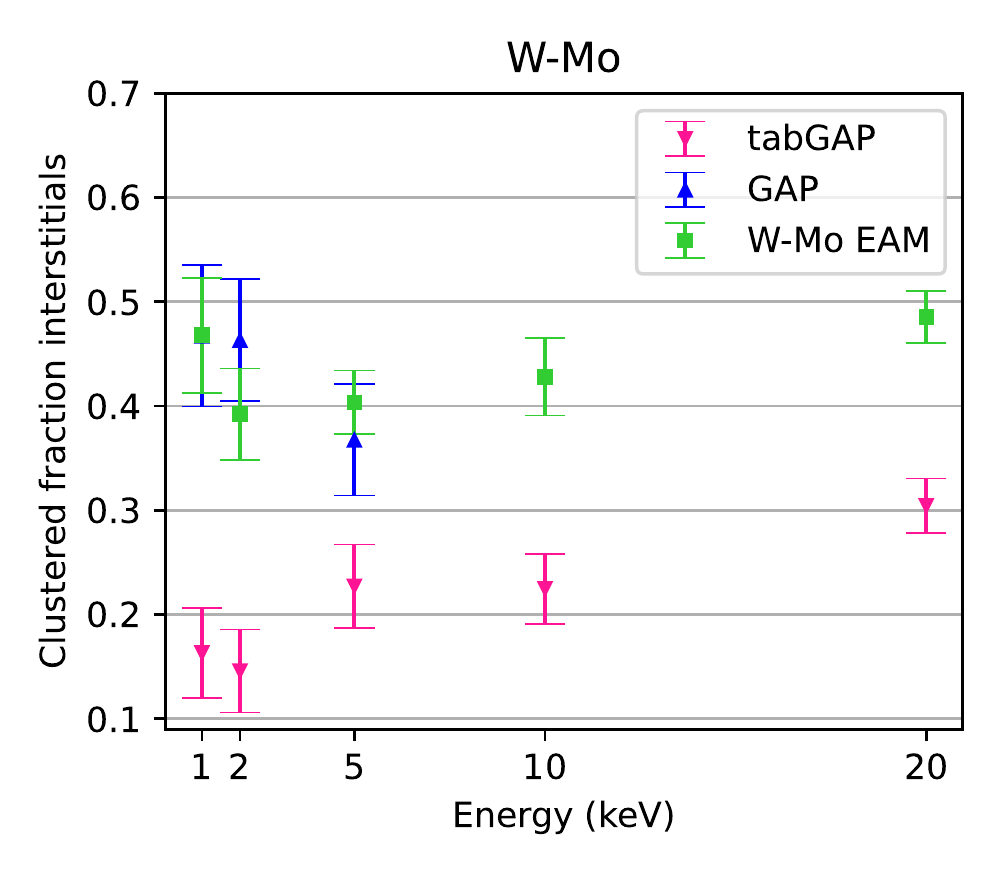}
        \caption{}
        \label{fig:cfrac_W-Mo_int}
    \end{subfigure}
    \begin{subfigure}{\columnwidth}
        \includegraphics[width=\linewidth]{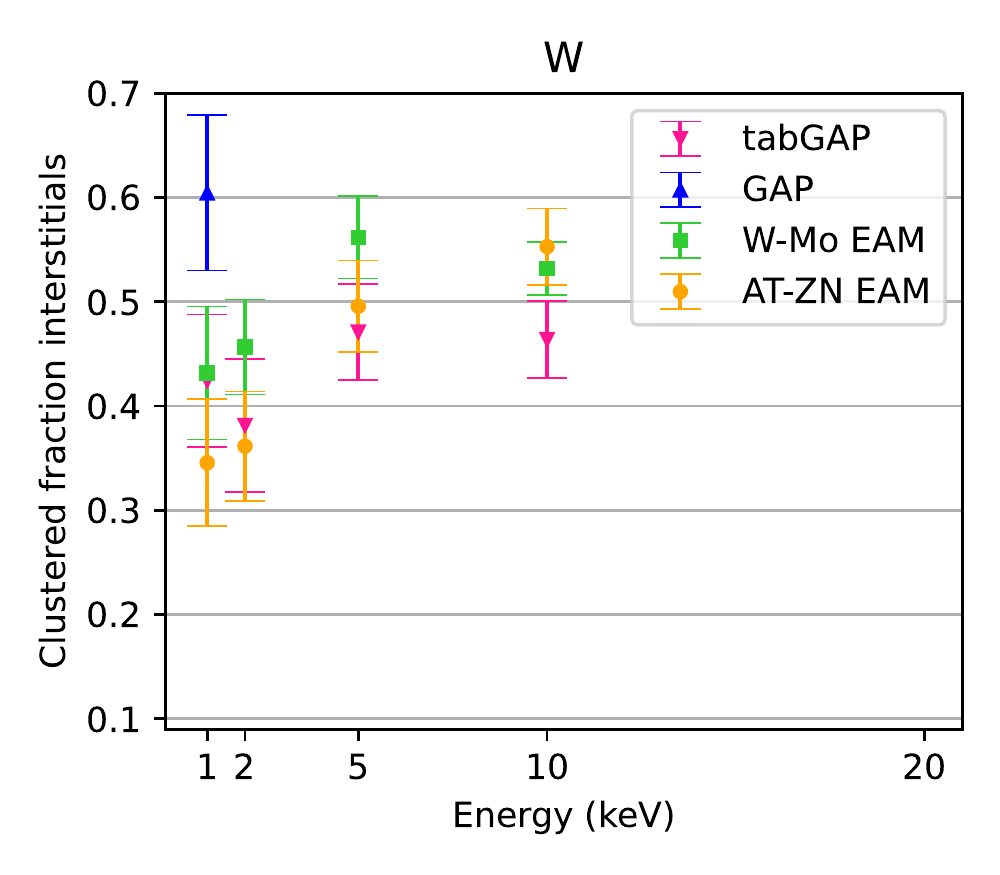}
        \caption{}
        \label{fig:cfrac_W_int}
    \end{subfigure}
    \\
    \begin{subfigure}{\columnwidth}
        \includegraphics[width=\linewidth]{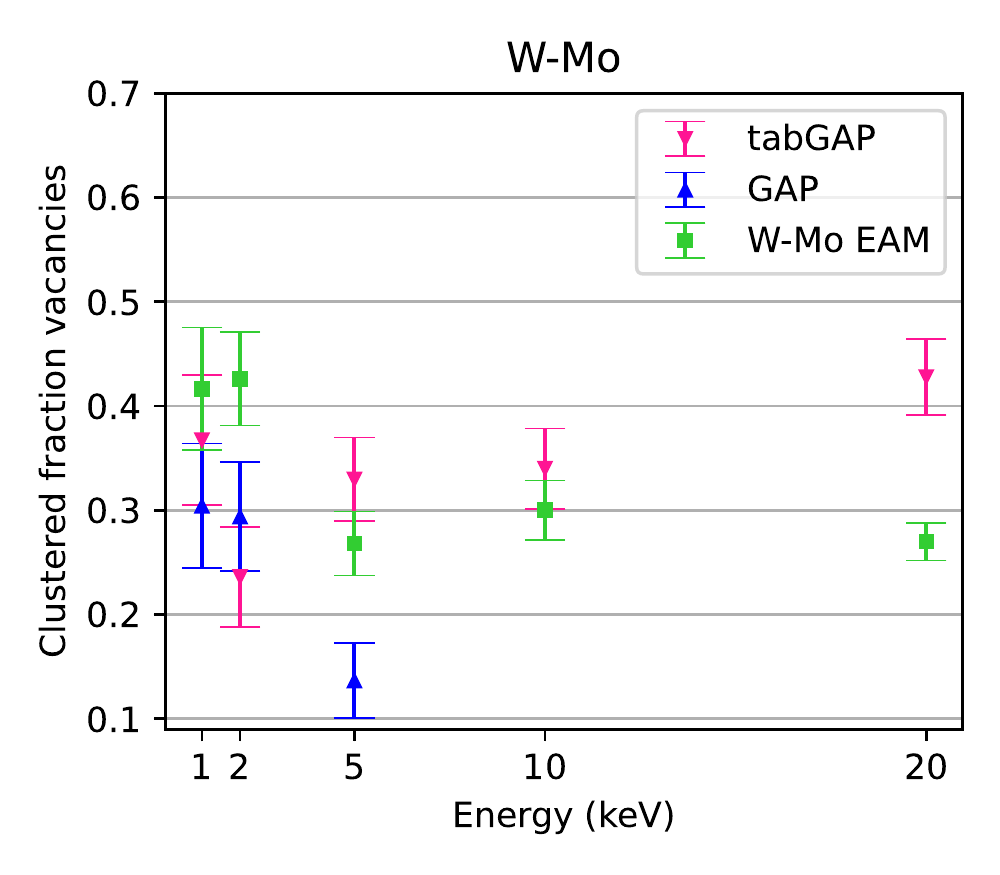}
        \caption{}
        \label{fig:cfrac_W-Mo_vac}
    \end{subfigure}
    \begin{subfigure}{\columnwidth}
        \includegraphics[width=\linewidth]{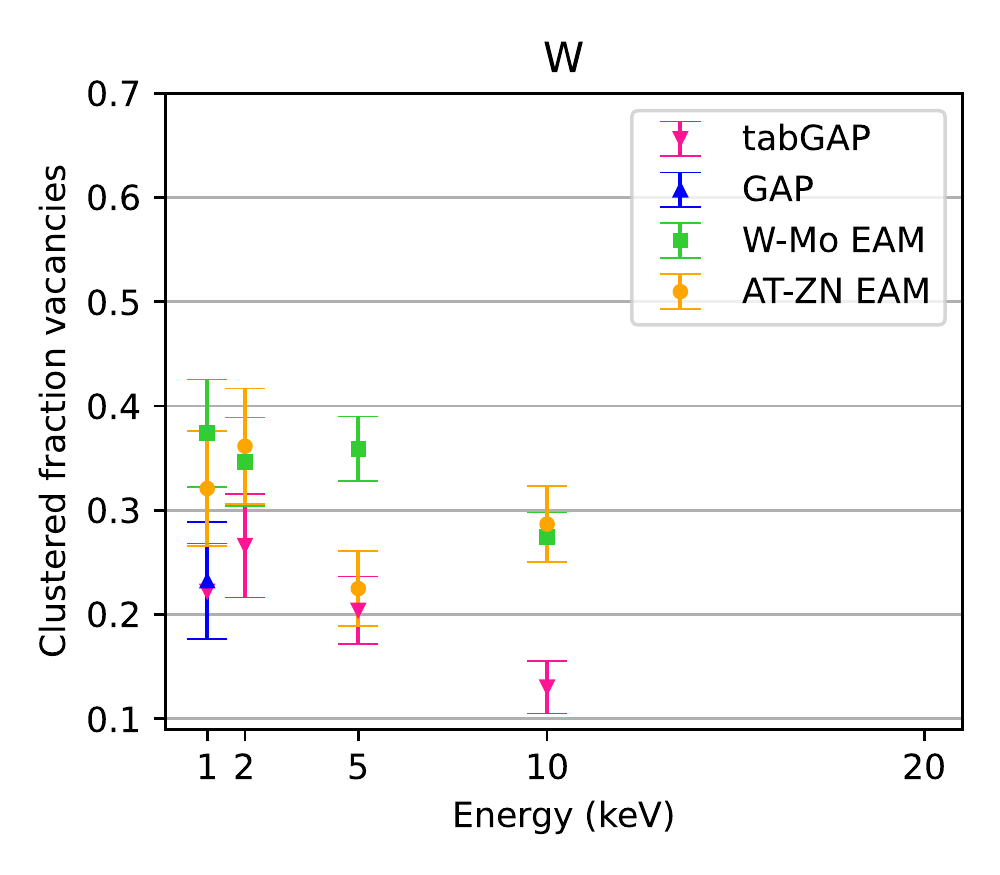}
        \caption{}
        \label{fig:cfrac_W_vac}
    \end{subfigure}
    \caption{Clustered fraction of defects. 
    The clustered fraction is computed as shown in Eq.~\ref{eq:cfrac}. Due to the clustered fraction not being defined for simulations with zero defects, only the simulations with non-zero defects are included in the  standard error. The vertical lines indicate the standard error. }
    
    \label{fig:cfrac}
\end{figure*}

The clustered fraction of defects is a quantity that allows us to analyse the clustering efficiency of the formed defects in a given potential. It is evaluated as follows:
\begin{equation}\label{eq:cfrac}
    \frac{N_{\text{tot}}-N_{\text{c}}}{N_{\text{tot}}}\, ,
\end{equation}
where $N_{\text{c}}$ is the number of defects, vacancies or interstitials, bound into clusters with a size greater than 1, and $N_{\text{tot}}$ is the total number of defects of the corresponding type.
This quantity is shown for W-Mo and W in Fig.~\ref{fig:cfrac}.

The cases with zero defects are excluded from this analysis because the clustered fraction is not defined in such cases.
Doing so does not affect the analysed quantity.

We see that the clustered fraction in tabGAP follows similar behaviour to that obtained with both EAM potentials. However, the clustered fraction for interstitials in W-Mo by tabGAP is somewhat lower compared to the W--Mo-EAM potential.
In pure W, the interstitial clustered fraction is quite similar for the EAMs and tabGAP, given the standard errors, whilst GAP resulted in more efficient clustering of interstitials.

In the case of vacancies, tabGAP predicted similar clustering in both W and W-Mo as GAP, with the only noticeable difference between the results being at 5~keV.
In general, we note that the tabGAP prediction of the interstitial clustering is less consistent with that of GAP, at least, within the statistical uncertainty available in the present work. This can be explained by the smaller training dataset for the W-Mo pair within the 5-element tabGAP potential.

The results of tabGAP imply that interstitials in W have a substantially higher tendency to form clusters than in W-Mo.
Surprisingly, both W--Mo-EAM and GAP predict a rather similar tendency for clustering, although, in all three potentials,  we see that the interstitials in W cluster more efficiently than in W-Mo. 
This is reasonable, given that interstitials are more mobile in W, and can therefore form clusters more swiftly than in W-Mo.
In the case of vacancies, only tabGAP and GAP reliably predict that vacancies are less clustered in W, as discussed above.

\subsection{Dislocation loops}

\begin{figure*}
    \begin{subfigure}{.7\columnwidth}
        \includegraphics[width=\linewidth]{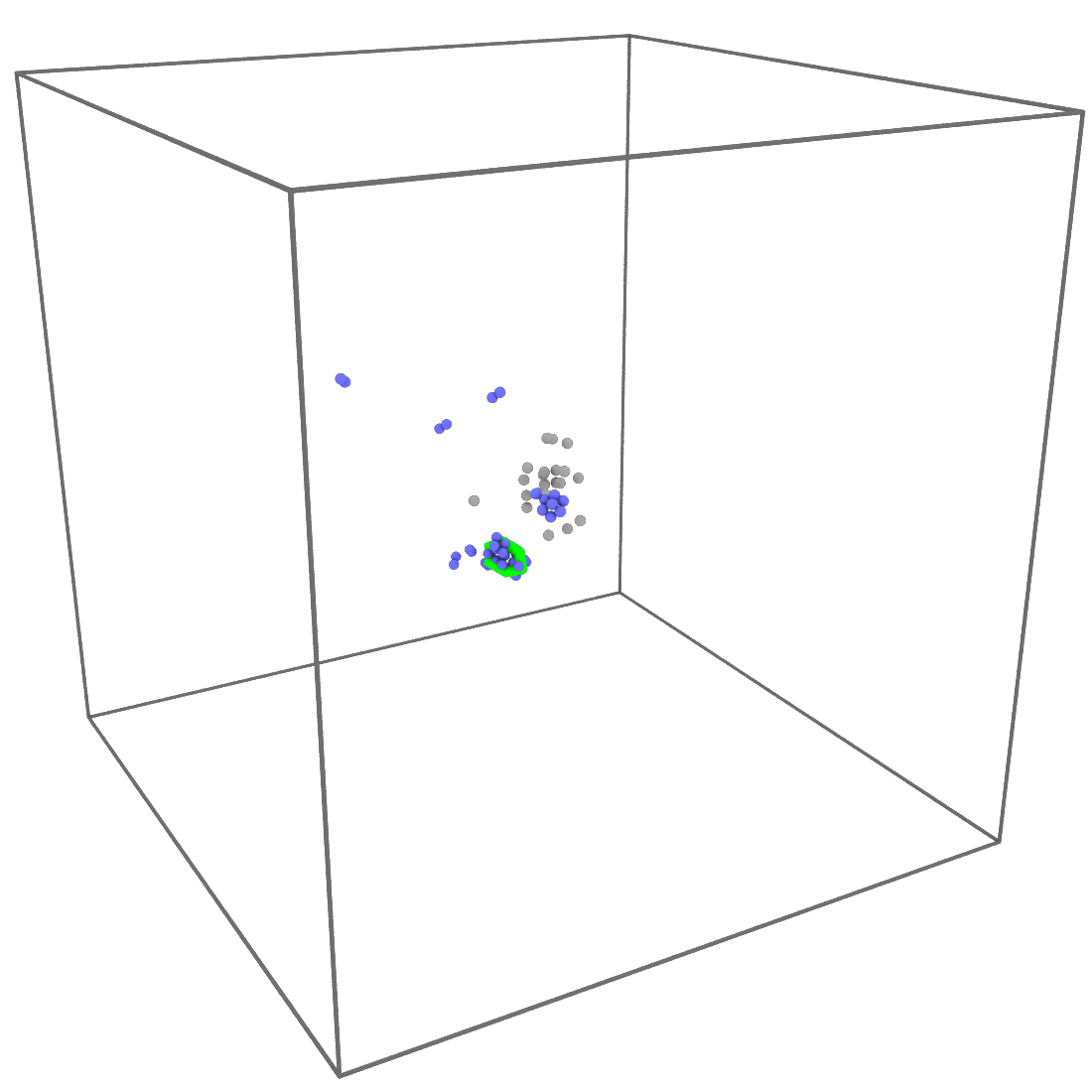}
        \caption{Pure W (AT-ZN EAM) 10~keV}
        \label{fig:AT_ZN_10_keV_dislocation_W}
    \end{subfigure}
    \hspace{2cm}
        \begin{subfigure}{.7\columnwidth}
        \includegraphics[width=\linewidth]{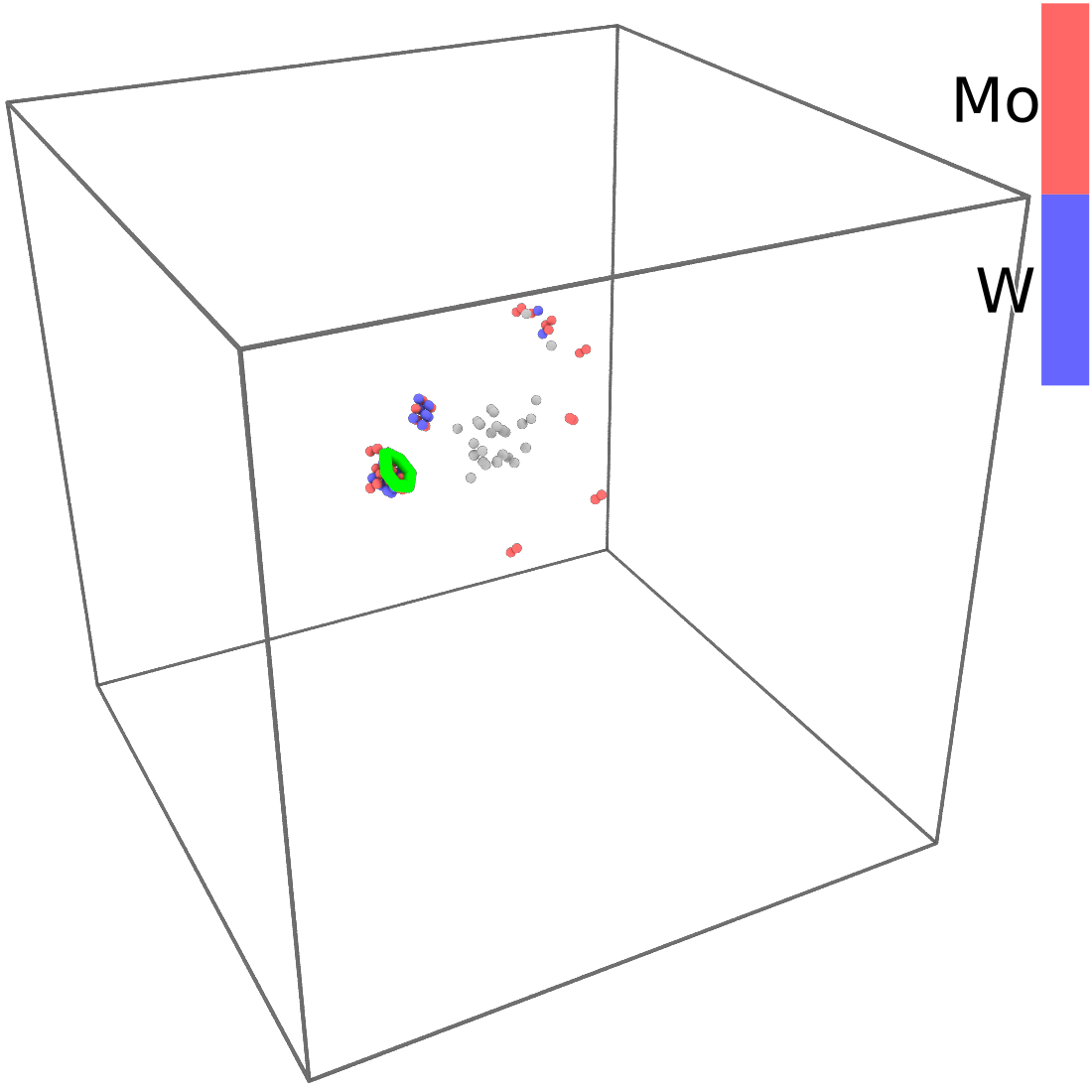}
        \caption{W--Mo (W--Mo-EAM) 10~keV}
        \label{fig:EAM_10_keV_dislocation}
    \end{subfigure}
     \\
    \begin{subfigure}{.7\columnwidth}
        \includegraphics[width=\linewidth]{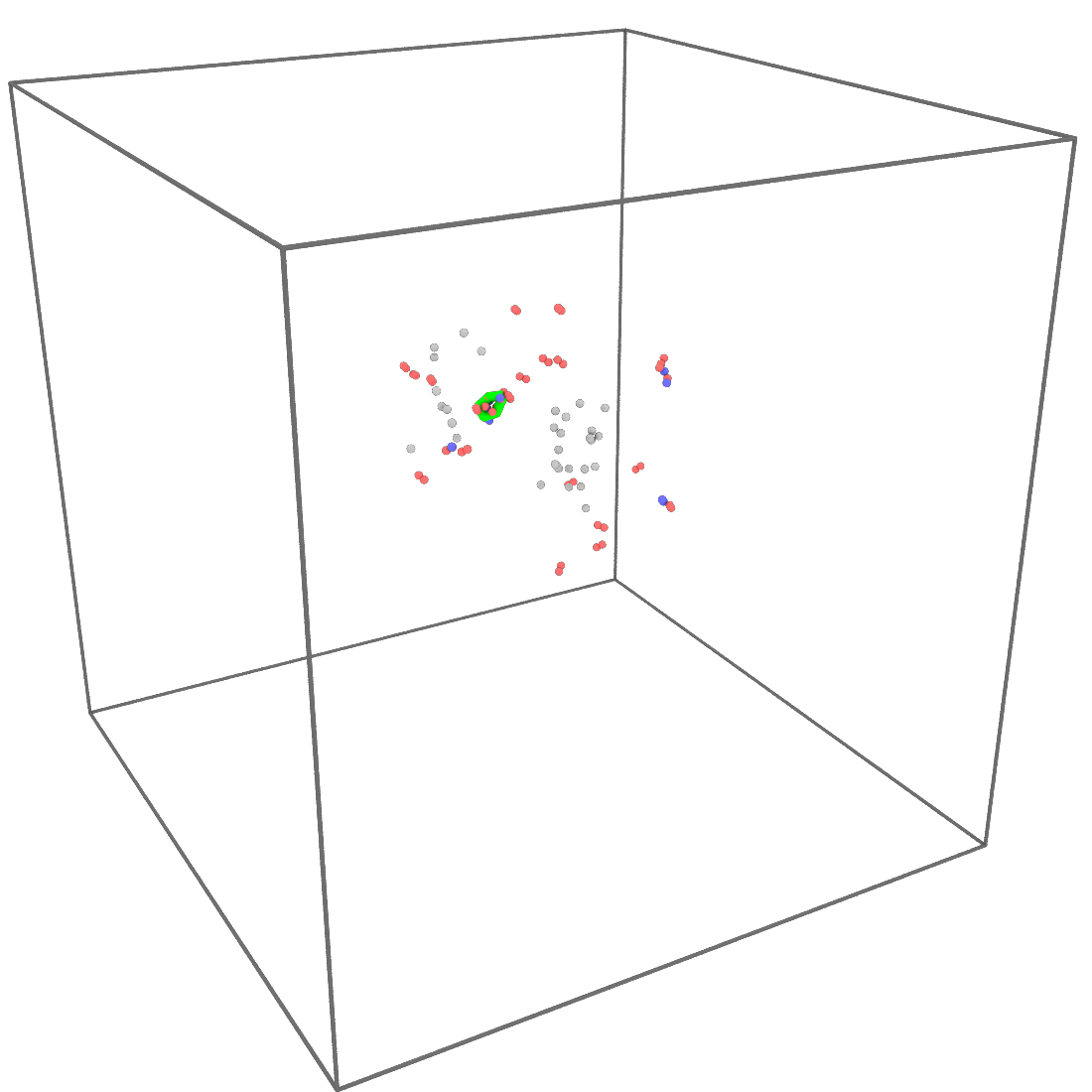}
        \caption{W--Mo (W--Mo-EAM) 20~keV}
        \label{fig:EAM_20_keV_dislocatio}
      \end{subfigure}
    \hspace{2cm}
    \centering
    \begin{subfigure}{.7\columnwidth}
        \includegraphics[width=\linewidth]{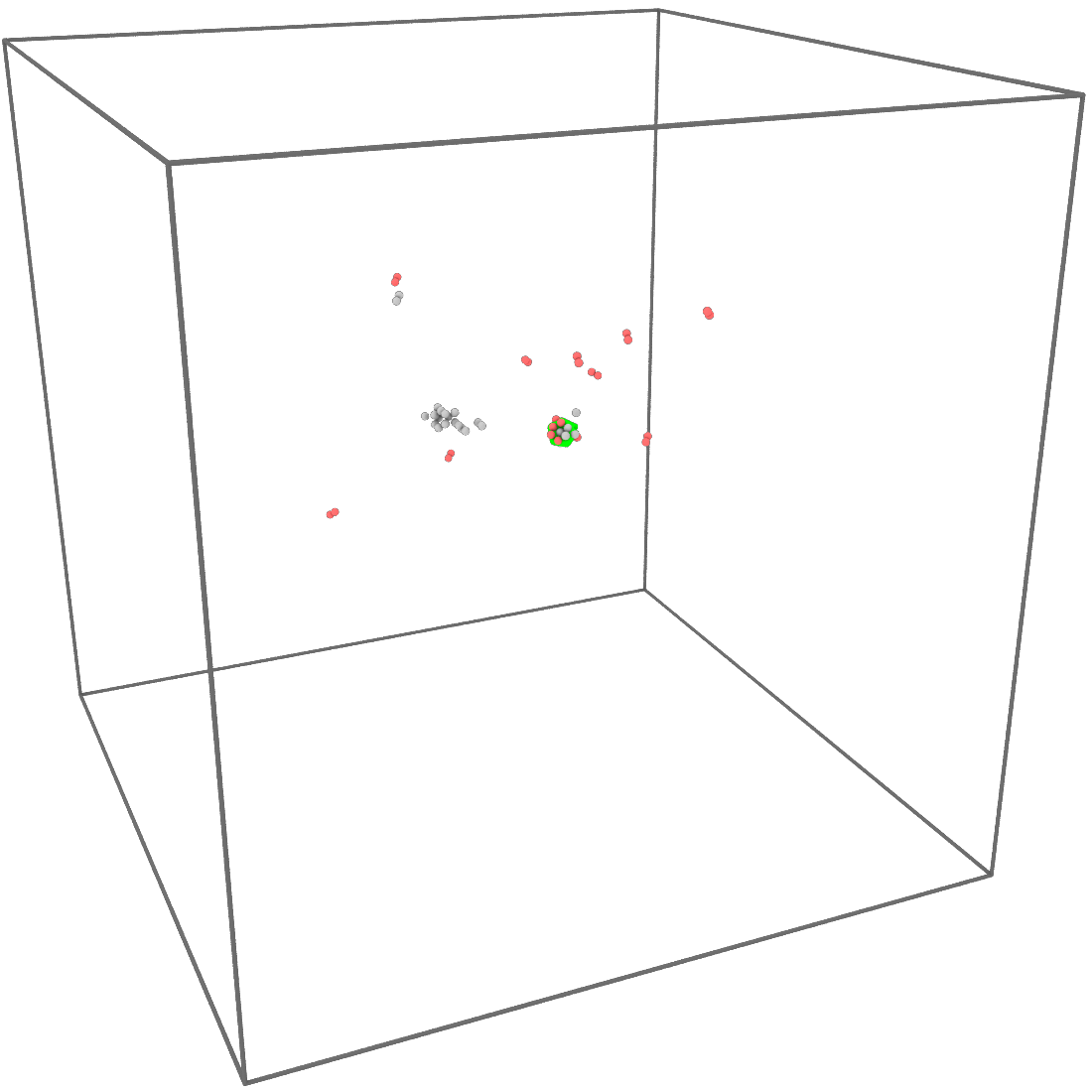}
        \caption{W--Mo (tabGAP) 20~keV}
        \label{fig:tabGAP_20_keV_dislocation}
    \end{subfigure}
    \caption{Defects at the end of the simulations. The grey particles are vacancies, the coloured particles are interstitials (Mo is red, W is blue), and $1/2~\left\langle 111 \right\rangle$ interstitial loops are shown as green lines.}
    \label{fig:dislocations}
\end{figure*}

The energetically most stable dislocation loops in W are those with Burgers vectors of $ 1/2~\left\langle 111 \right\rangle$~\cite{W_dislocation_loops_experimental, W_dislocation_loops_DFT}.
In all W-Mo cascades, there were only three cases, of dislocations identified by the DXA algorithm in \textsc{ovito}, whereas pure W only had one case in an AT-ZN EAM simulation. These dislocations were small loops of the  interstitial type, formed in the  10- and 20-keV cascades (10 keV in the case of W). The observed dislocations were all $1/2~\left\langle 111 \right\rangle$, as shown in Fig.~\ref{fig:dislocations}. 

\subsection{Performance}
It is imperative to discuss the difference in performance between the potentials since it was the motivation for developing tabGAP. 
For example, 100-ps, 5-keV tabGAP simulations using 12 processing cores were completed in less than a day, whereas GAP required a run-time of three days to attain 70~ps simulated time using 1~000 cores.\\
\\
It is worth noting that the tabGAP framework has been further developed after the present simulations using tabGAP had been performed. The new version developed in Ref.~\cite{byggmastar_simple_2022} has optimised code and cut-off radii, and includes an EAM-like energy contribution, which makes it both more accurate and faster than the tabGAP used in this study. In light of this, the performance of the newer tabGAP, called here enhanced tabGAP (e-tabGAP), was tested in addition to the four potentials used in this study. For more details and benchmarks of the e-tabGAP, we refer to Ref.~\cite{byggmastar_simple_2022}.
 
The performance of the potentials was tested by running $N\,P\,T$ simulations in 31~250~-atom cells
These simulations were run for 2~000, 3-fs time-steps, using 30 central processing unit cores. The results are provided in Table~\ref{tab:performance}.
\begin{table}[h]
\begin{ruledtabular}
\centering
\caption{Performances of the potentials. Here, e-tabGAP denotes the newer version of tabGAP~\cite{byggmastar_simple_2022}; $t_{\text{EoM}}$ denotes the time it took to evaluate the equation of motion of a single atom; $t_{\text{loop}}$ denotes the loop-time given by LAMMPS, which is the total wall-clock time elapsed from the start to the evaluation of the last equation of motion; $s$ is the performance in units of GAP, i.e. how many times faster a given potential is than GAP.
}
\label{tab:performance}			
\setlength{\tabcolsep}{0pt}  
\renewcommand{\arraystretch}{1.2}
\begin{tabular}{cccc}
Potential & $t_{\text{EoM}} \left[\mu\text{s}\right]$ & $t_{\text{loop}}\,  \left[\text{\text{h}} \right]$ & $s \,\left[ \text{GAP} \right]$ \\
\midrule 
AT-ZN EAM &  1.7   & 0.001 & 49~000\\
W--Mo-EAM &  4.4   & 0.003 & 19~000\\
e-tabGAP  &  50    & 0.03  & 1~700 \\
tabGAP    &  360   &  0.3  & 230 \\
   GAP    & 83~000 &  48   &  1 
\end{tabular}
\end{ruledtabular}
\end{table}
From Table~\ref{tab:performance}, it is evident how slow GAP is compared to the other potentials. The tabGAP used in this work is roughly two orders of magnitude faster than GAP, and two orders slower than the EAMs. With the newer version, e-tabGAP, the speed-up is three orders of magnitude to GAP, and only \textit{one} order of magnitude slower than the EAMs. The primary sources of discrepancy in the EAM performances are the larger cut-off radii used in the W--Mo-EAM as opposed to the AT-ZN variant.

To put the difference in the performances of GAP and e-tabGAP into perspective, let one consider the following example: given the same computational resources and the same task, a job that would take e-tabGAP \textit{three days}, would take GAP closer to 14 \textit{years}.

\section{Summary of observations}
For clarity, we here summarise the observations discussed in the previous sections. The comparison between tabGAP and GAP can be summarised as follows:

\begin{enumerate}
    \item TabGAP was found to be two orders of magnitude faster than GAP, and two orders of magnitude slower than the EAM potentials. The newer version of tabGAP (optimised code and cut-off radii) is three orders faster than GAP, and one order slower than the EAMs.
    
    \item The number of surviving Frenkel pairs in tabGAP was found to be close to GAP, albeit always slightly higher, within the uncertainties given by the standard error of the mean.
    
    \item TabGAP and GAP produced similar defect-clustering, within the standard error bars, although there is some difference in the number of specific cluster sizes between the two potentials.
    
    \item We also found that, overall, the fraction of interstitial atoms bound into clusters was smaller in tabGAP than in GAP. The cause for this discrepancy may lie in the smaller training data for tabGAP. 
    
\end{enumerate}
The differences between 50-50 W-Mo alloy and pure W in the primary radiation damage can be summarised as:
\begin{enumerate}
    \item Interstitials at a given temperature in W-Mo were found to be substantially less mobile than in W.
    
    \item All interstitials in W-Mo and W were  split-interstitials.
    
    \item Mo-Mo interstitials were the predominant interstitials in W-Mo.
    
    \item Interstitial clusters in W were larger than in W-Mo. This is likely a result of the superior mobility of interstitials in W allowing for more rapid clustering, as opposed to W-Mo.
    
    \item Small vacancy clusters in W-Mo were found to be more abundant than in W, according to tabGAP and GAP, whereas the W--Mo-EAM the opposite, albeit to a lesser extent. Our DFT results show that divacancies in W-Mo are almost as unstable as in W, which none of the potentials can reproduce. This could imply that smaller vacancy clusters are as abundant in the 50-50 alloy as in pure W. 

    \item The 50-50 W-Mo had on average the same number of defects as pure W, which implies that the presence of Mo has no significant effect on the cascade dynamics.
    
    \item However, we noticed slightly more efficient recombination of defects in the 50-50 W-Mo alloy, since there were several cases where the defects created in cascades fully recombined. This behaviour was not observed in pure W.
    Additionally, W-Mo was observed to recombine a greater fraction of defects produced during the early phase of the cascades.

\end{enumerate}
\section{Conclusions}

The aim of this study was to analyse the benefits and possible drawbacks of a more efficient version of the machine-learning potential GAP, the so-called tabGAP. 
In this study, we report the differences and similarities between pure W, and W-Mo (50:50) alloy with respect to the primary radiation damage as predicted by three potentials:
tabGAP, GAP, and EAM. In W-Mo, the main difference between EAM and (tab)GAP is the number of surviving defects, which is significantly higher in the EAM potential. However, in pure W, the well-established AT-ZN EAM potential produces similar numbers of defects and clustering statistics to tabGAP, which are also fairly similar to the available predictions made by GAP and much lower than the values predicted by the W--Mo-EAM potential.

We conclude that, overall, tabGAP produces similar results to GAP in cascade simulations in a random binary alloy, while being two orders of magnitude faster. This makes tabGAP a promising machine-learned potential for accurate modelling of low- and high-dose radiation damage in multicomponent alloys.

\section{Acknowledgements}

We are grateful for funding from the Academy of Finland project
HEADFORE (grant no. 333225).
This work has been partly carried out within the framework of the EUROfusion Consortium, funded by the European Union via the
Euratom Research and Training Programme (Grant Agreement No 101052200 — EUROfusion). Views and opinions expressed are however those of the author(s) only and do not necessarily reflect those of the European Union or the European Commission.
 Neither the European Union nor the European Commission can be held responsible for them.
The authors wish to thank the Finnish Grid and Cloud Infrastructure (FGCI) 
(persistent identifier urn:nbn:fi:research-infras-2016072533) for supporting this project with computational and data storage resources. 

\bibliography{main_bibliography.bib}

\newpage
\section{Appendix}\label{sec:appendix}
\subsection{Time-integration error}
\begin{figure} 
     \begin{subfigure}{\columnwidth}
         \includegraphics[width=.8\linewidth]{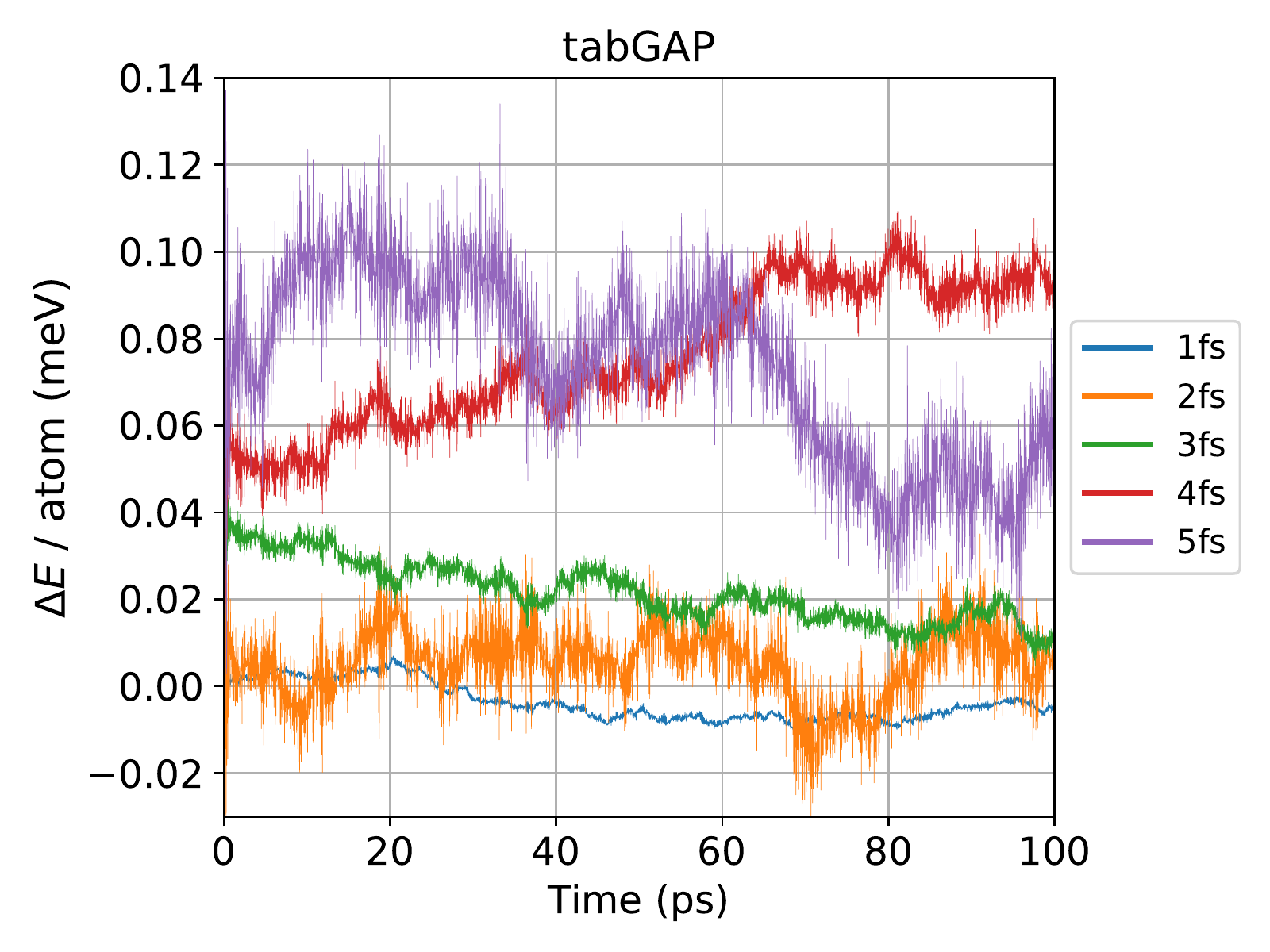}
         \caption{}
         \label{fig:NVE_tabGAP}
     \end{subfigure}\\
     \begin{subfigure}{1\columnwidth}
         \includegraphics[width=.8\linewidth]{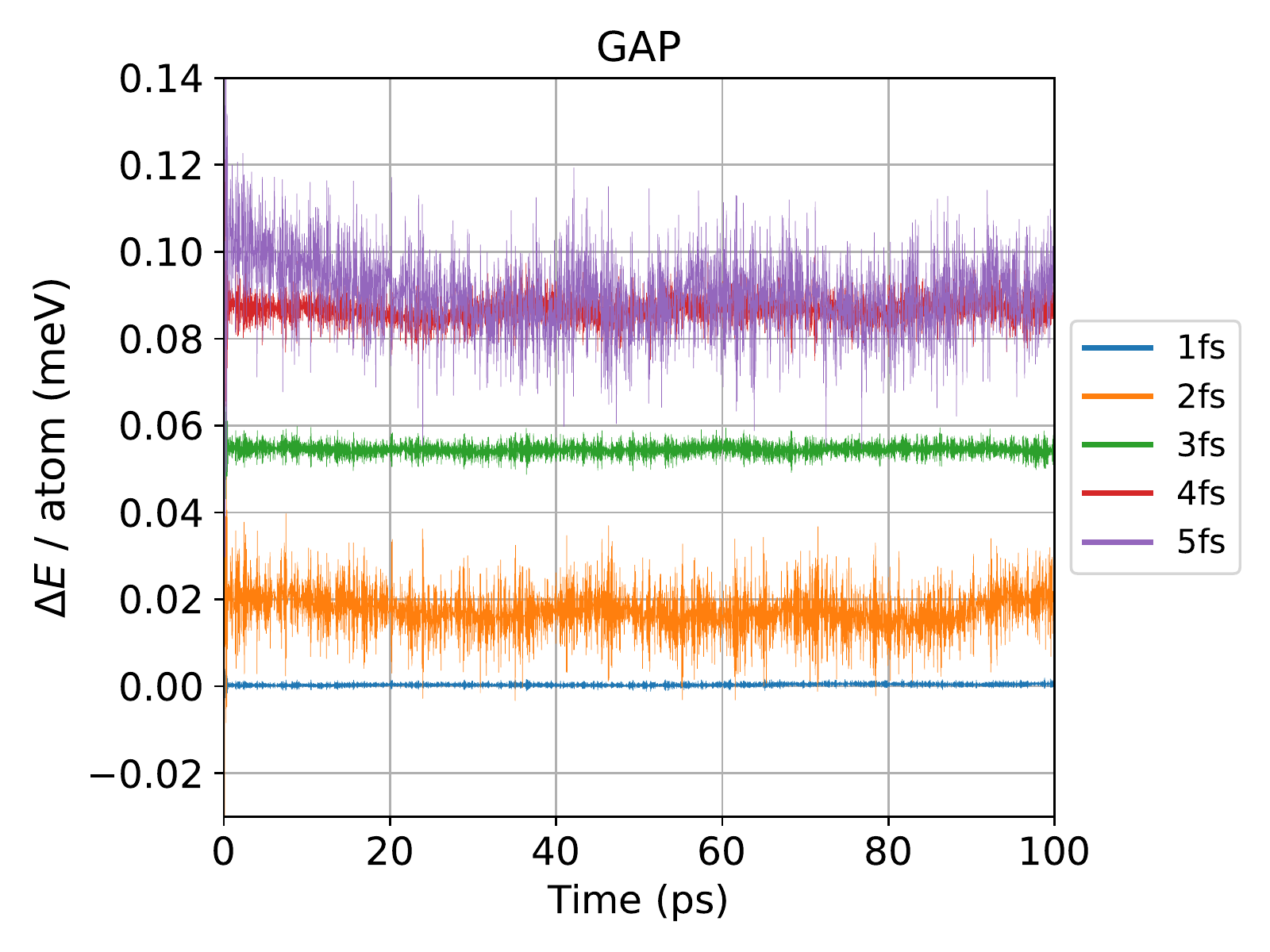}
         \caption{}
         \label{fig:NVE_GAP}
     \end{subfigure}\\
     \begin{subfigure}{\columnwidth}
         \centering
         \includegraphics[width=.8\linewidth]{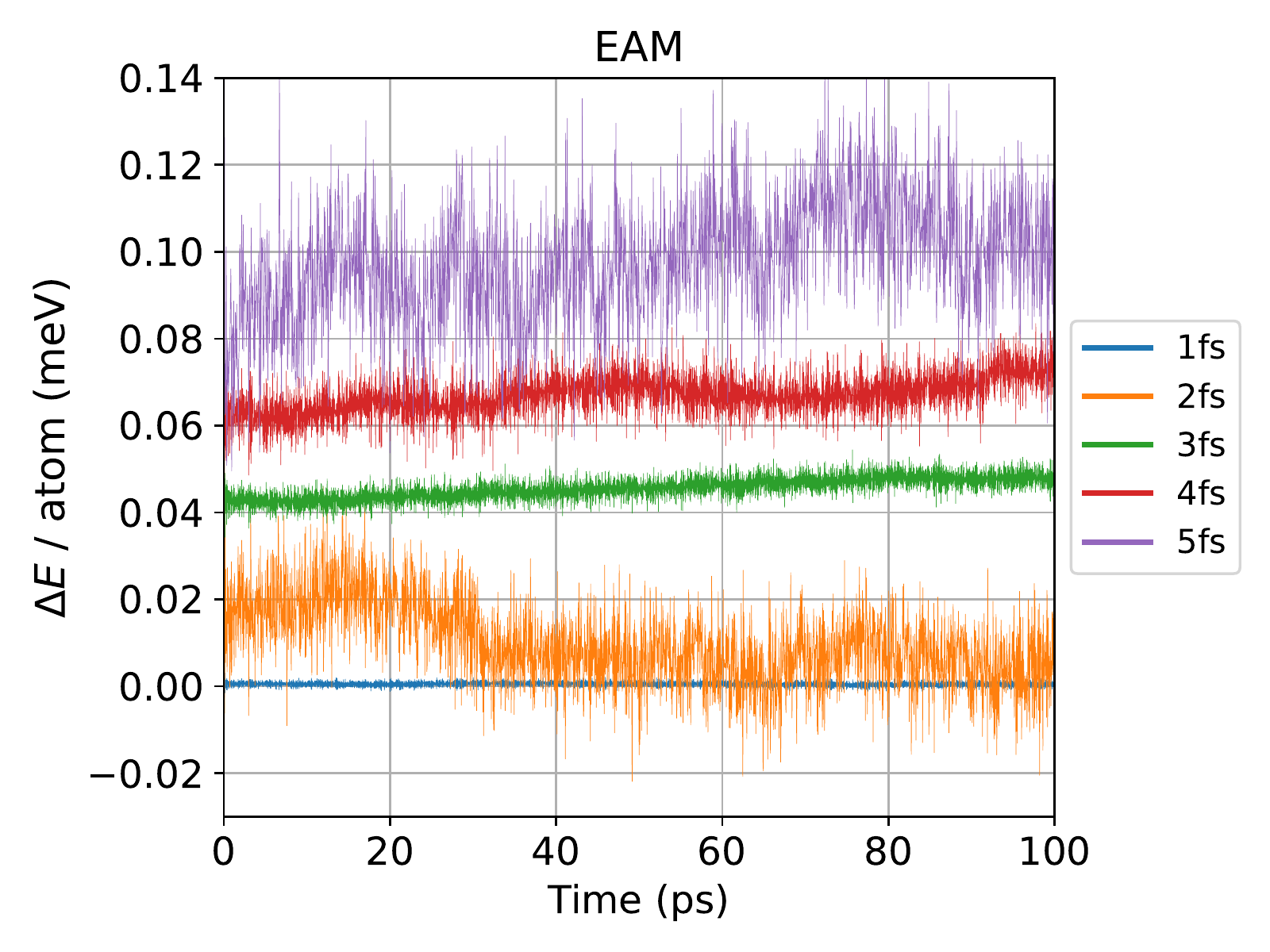}
         \caption{}
         \label{fig:NVE_EAM}
\end{subfigure}
        \caption{The y-axis shows total-energy variations per atom in a W-Mo $NVE$ ensemble of 1~024 atoms. The y-axis values have been shifted for clarity, but the magnitudes of relative changes therein are unchanged. The energy variations are computed by subtracting each energy value from a fixed value and dividing it by the number of atoms in the cell.}
        \label{fig:NVE}
\end{figure}

\begin{minipage}{\columnwidth}
Here we compare the time-integration error between the three potentials. To test this, we ran test simulations, using the velocity Verlet algorithm, in cells comprising $1~024$ atoms, that were not connected to thermostats or barostats, making them $NVE$ ensembles; ensembles where the total energy should stay constant.
In Fig.~\ref{fig:NVE}, one can see the results from simulations for all of the potentials for varying values of time-step, using the aforementioned cell at a temperature of $500~\text{K}$; the flatter the line, the better. Fluctuations of total energy in an $NVE$ ensemble are due to time-integration error, caused by having a non-zero time-step.
 \\
 \\
 Interestingly, tabGAP shows erratic variation in total energy (Fig.~\ref{fig:NVE_tabGAP}), whereas EAM and GAP show more consistency in the pattern of the variation. The erratic variation of tabGAP could be caused by interpolation error. Even so, the largest fluctuation per atom ($5-\text{fs}$ time-step) is only $\approx 0.15~\text{meV}$, whereas for GAP and EAM respectively, these are $\approx 0.06~\text{meV}$ and $\approx 0.08~\text{meV}$. The average kinetic energy of an atom in these simulations is $\frac{3}{2}~k_{\text{B}}~500~\text{K}~\approx~65~\text{meV}$. Therefore, changes in the energy of an atom caused by tabGAP are completely masked by thermal vibrations and are thus insignificant.
 \end{minipage}
 \newpage
\end{document}